\documentclass[aps,prb,twocolumn,floats,epsfig]{revtex4}
\usepackage[iso-8859-1]{inputenx}
\usepackage{amssymb}
\usepackage{amsbsy}
\usepackage{amsmath}
\usepackage{epsfig}
\usepackage{color}
\usepackage{float}
\usepackage[colorlinks,linktocpage,bookmarks=false,citecolor=blue,linkcolor=red,urlcolor=blue]{hyperref}

\newcommand{\bib}{\bibitem}
\newcommand{\beq}{\begin{equation}}
\newcommand{\eeq}{\end{equation}}
\newcommand{\bea}{\begin{eqnarray}}
\newcommand{\eea}{\end{eqnarray}}
\newcommand{\al}{\alpha}
\newcommand{\de}{\delta}
\newcommand{\De}{\Delta}
\newcommand{\ep}{\epsilon}
\newcommand{\ga}{\gamma}

\newcommand{\si}{\sigma}

\newcommand{\om}{\omega}

\newcommand{\bra}[1]{\langle#1|}
\newcommand{\ket}[1]{|#1\rangle}

\newcommand{\non}{\nonumber}

\usepackage{xcolor}

\begin{document}

\title{Dynamical relaxation of correlators in periodically driven integrable
quantum systems}

\author{Sreemayee Aditya$^1$, Sutapa Samanta$^2$, Arnab Sen$^2$,
K. Sengupta$^2$ and Diptiman Sen$^1$}

\affiliation{$^1$Centre for High Energy Physics, Indian Institute
of Science, Bengaluru 560012, India \\
$^2$School of Physical Sciences, Indian Association for the
Cultivation of Science, Jadavpur, Kolkata 700032, India}

\date{\today}

\begin{abstract}

We show that the correlation functions of a class of periodically
driven integrable closed quantum systems approach their steady state
value as $n^{-(\alpha+1)/\beta}$, where $n$ is the number of drive
cycles and $\alpha$ and $\beta$ denote positive integers. We find
that generically $\beta=2$ within a dynamical phase characterized by
a fixed $\alpha$; however, its value can change to $\beta=3$ or
$\beta=4$ either at critical drive frequencies separating two
dynamical phases or at special points within a phase. We show that
such decays are realized in both driven Su-Schrieffer-Heeger (SSH)
and one-dimensional (1D) transverse field Ising models, discuss the
role of symmetries of the Floquet spectrum in determining $\beta$,
and chart out the values of $\alpha$ and $\beta$ realized in these
models. We analyze the SSH model for a continuous drive protocol
using a Floquet perturbation theory which provides analytical
insight into the behavior of the correlation functions in terms of
its Floquet Hamiltonian. This is supplemented by an exact numerical
study of a similar behavior for the 1D Ising model driven by a
square pulse protocol. For both models, we find a crossover
timescale $n_c$ which diverges at the transition. We also unravel a
long-time oscillatory behavior of the correlators when the critical
drive frequency, $\omega_c$, is approached from below ($\omega <
\omega_c$). We tie such behavior to the presence of multiple
stationary points in the Floquet spectrum of these models and
provide an analytic expression for the time period of these
oscillations.

\end{abstract}

\maketitle

\section{Introduction}

Non-equilibrium dynamics of closed quantum systems has been the
subject of intense research activity in the recent past
\cite{rev1,rev2,rev3,rev4,rev5,rev6,rev7}. Theoretical studies
on the subject focussed initially on quench
\cite{quench1,quench2,quench3} and ramp
\cite{ramp1,ramp2,ramp3,ramp4,ramp5,ramp6} protocols. However,
recently the focus in the field has shifted to periodically driven
systems \cite{rev5, rev6, rev7}. More recently quasi-periodic and
aperiodically driven systems have also been studied
\cite{quasi1,quasi2,quasi3,ap1,ap2}. The experimental signatures of
such dynamics have been investigated in the context of ultracold atoms in
optical lattices \cite{exp1,exp2,exp3,exp4,exp5}.

Quantum systems driven out of equilibrium via a periodic protocol
host several phenomena which are not seen in those driven by a quench
or a ramp. These include the generation of drive-induced topological
states of matter \cite{topo1,topo2,topo3}, realization of Floquet
time crystals \cite{tc1,tc2,tc3}, and phenomena such as
dynamical localization \cite{dynloc1,dynloc2,dynloc3,dynloc4},
dynamical freezing \cite{dynfreez1,dynfreez2,dynfreez3}, and drive-induced
tuning of ergodicity\cite{erg1,erg2}. These
properties of periodically driven systems, having a time period $T$,
are most easily understood from their Floquet Hamiltonian $H_F$
which is related to their unitary evolution operator $U$ via the
relation $U(T,0)= \exp[-i H_F T/\hbar]$ \cite{rev7}.

The presence of dynamical transitions constitutes yet another
interesting phenomenon in periodically driven closed quantum system
\cite{dyntran1,dyntran2, dyntran3, asen1, dyntran5, dyntran6}. Such
transitions can be categorized into two distinct classes. The first
involves non-analyticities of the return probability of its wave
function; these non-analyticities show up as cusps in Loschmidt
echoes \cite{dyntran1}. Such transitions can be related to Fischer
zeroes of the complex partition function of the driven system
\cite{dyntran1,dyntran2}. In contrast, the second class of
transitions constitutes a change in the long-time behavior of the
correlation functions of a periodically driven integrable quantum
system as a function of the drive frequency \cite{asen1, dyntran5}.
Such a transition results from a change in the extrema of the
Floquet Hamiltonian $H_F$ as a function of the drive parameters; the
signature of such transitions can be deciphered from the study of
local correlation functions of such models \cite{asen1,dyntran5}.
The study of such transitions has also been extended to integrable
models with long-ranged interactions \cite{dyntran5} and those
coupled to an external bath \cite{dyntran6}. The characteristics of
the correlation function in the two dynamical phases across the
transition have been studied in detail. It was shown that for a
$d$-dimensional integrable system after $n$ drive cycles and for
large $n$, these correlators decay as $n^{-(d+2)/2}$ in the
high-frequency regime and as $n^{-d/2}$ in the low-frequency regime.
However, the behavior of the system at a dynamical critical point
and its vicinity has not been studied previously.

In this work, we study the properties of correlation functions
for general driven 1D integrable quantum systems which
have a simple representation in terms of free fermions. Our analysis
holds for several 1D spin systems such as the Ising model in a
transverse field, the $XY$ model, and the 1D Kitaev chain. All of
these models allow for a simple fermionic representation via a
Jordan-Wigner transformation leading to a quadratic, exactly
solvable Hamiltonian \cite{subir1}. In addition, it is also
applicable to charge- or spin-density wave systems described by the
SSH model \cite{sshref}.

The central points that emerge from such a study are as follows.
First, we show that all local fermionic correlation functions of
such driven models decay to their steady state value according to
the relation
\begin{eqnarray}
C_{x}(nT) \sim n^{-(\alpha+1)/\beta}, \label{mainr} \end{eqnarray}
where $\alpha$ and $\beta$ are positive integers and $x$ indicates
the spatial coordinate. We note that only the case of $\beta=2$ and
$\alpha=0,~2$ has been discussed in earlier studies
\cite{asen1,dyntran5,dyntran6}; these are reproduced as special
cases of the general result given by Eq.\ \eqref{mainr}. We show
that such a result is tied to the stationary point structure and
symmetry properties of the Floquet spectrum of the system. We
identify the condition for the existence of anomalous powers ($\beta
\ne 2$) for the driven system and estimate a crossover scale, $n_c$,
after which the system is expected to deviate from the anomalous
($\beta \ne 2$) scaling towards the generic ($\beta=2$) one. This
crossover scale diverges at specific points in the parameter space
of the driven system; we chart out the condition for the realization
of such points in terms of its Floquet spectrum. Second, we provide
specific example of such decay with $\beta \ne 2$ in the context of
simple models. To this end, we study the driven SSH model using a
continuous drive protocol. We find the realization of decay
exponents $-1/3$ corresponding to $\beta=3$ and $\alpha=0$. We
analytically calculate the corresponding Floquet Hamiltonian within
a Floquet perturbation theory (FPT) \cite{rev7,fpt1,fpt2} which
provides insight into the structure of the Floquet spectrum and the
correlation functions of the model. Such analytical results are
shown to match closely with exact numerical studies. Third, we
identify a long-time coherent oscillation of the correlation
function of such models when the drive frequency is near to but less
than a critical drive frequency. We show that the oscillation is a
consequence of the presence of multiple stationary points in the
Floquet spectrum of the SSH model; consequently, it is absent at
drive frequencies higher than the critical frequencies. We provide
an analytic expression for the time period of the oscillation which
matches our numerical results. Fourth, we analyze the Ising model
driven by a square pulse protocol and show the existence of
anomalous decay exponents corresponding to $\beta=4$ at the first
dynamical transition. We provide a detailed analysis of the
crossover scale around this transition. Furthermore, we note that at
the reentrant transitions present in this model, the correlation
functions show a decay characterized by an exponent of $-1/3$ which
is similar to that in the SSH model. In addition, near the first
transition, we unravel the long-time oscillatory nature of the
correlation functions when the critical drive frequency is
approached from below (lower frequency); this feature is absent when
the transition is approached from above. We provide an explanation
of such a behavior using the properties of the Floquet spectrum of
the driven models. Finally, our analysis identifies a crossover
scale $n_c$ which diverges at the dynamical transition characterized
by the critical drive frequency $\omega_c$: $n_c \sim
|\omega-\omega_c|^{-\beta_0/(\beta_0-a_0)}$, where $a_0=1$ or $2$
depending on the symmetry of the model, and $\beta_0>a_0$
corresponds to the order of the second term in expansion of the
Floquet energy around the transition point. For $n>n_c$, the decay
of the correlation function follows a generic exponent corresponding
to $\beta=2$; below $n_c$, the decay is characterized by $\beta>2$.
We validate such a power-law divergence of $n_c$ from exact numerics
for both the Ising and the SSH model.

The plan of the rest of the paper is as follows. In Sec.\
\ref{genres}, we analyze the correlation functions of a driven
fermion model and provide a detailed derivation of Eq.\
\eqref{mainr}. This is followed, in Sec.\ \ref{ssh}, by a study of
the driven SSH model which provides concrete examples of the scaling
laws discussed. Next, in Sec.\ \ref{ising}, we study the scaling
behavior of the correlation functions of the periodically driven 1D
Ising model in a transverse field. Finally, we summarize our results
and conclude in Sec.\ \ref{diss}.

\section{General results}
\label{genres}

In this section, we shall discuss the general behavior of
correlation functions of periodically driven 1D integrable models.
In what follows, we shall consider a 1D integrable model whose
Hamiltonian is given by
\begin{eqnarray}
H &=& \sum_k \psi_k^{\dagger} H_k \psi_k, \quad H_k = {\vec \sigma}
\cdot \vec h(k,t), \label{hamdef} \end{eqnarray} where $\vec \sigma
= (\sigma_x, \sigma_y, \sigma_z)$ denotes the standard Pauli
matrices, and $\vec h(k,t)= (h_x(k,t), h_y(k,t), h_z(k,t))^T$ is the
Hamiltonian density in momentum space. The time-dependence of $\vec
h(k,t)$ is fixed by the drive; in this work, we shall consider the
case where the drive is characterized by a time period $T=
2\pi/\omega$, where $\omega$ is the drive frequency. In what
follows, we shall consider $\psi_k = (a_k, b_k)^T$ to be a
two-component fermionic field characterized by annihilation
operators $a_k$ and $b_k$. The exact nature of these operators
depend on the model and shall be discussed in detail in subsequent
sections for the SSH and the Ising models.

The unitary evolution operator for such systems can be expressed in
term of their Floquet Hamiltonian
\begin{eqnarray}
U(T,0) &= \prod_k U_k(T,0) = T_t e^{-i \int_0^T dt H(t)/\hbar} =
e^{-i H_F T/\hbar}, \non \\
\label{udef} \end{eqnarray}
where $H_F$ is the Floquet Hamiltonian of the system. Thus $U_k$ for
such models can be resolved in terms of the Floquet eigenvalues and
eigenvectors. Since $U_k(T,0)$ is a $2\times 2$ matrix, we find
\begin{eqnarray}
U_k(T,0) &=& \sum_{j=1,2} e^{ -i \epsilon_F^{(j)}(k) T/\hbar}
|n_j(k)\rangle \langle n_j(k)|, \label{evolop} \end{eqnarray}
where $\epsilon_F^{(j)}(k)$, for $j=1,2$, are the Floquet
eigenvalues and $|n_j(k)\rangle$ are the corresponding eigenvectors.

To compute the correlation functions for such a driven system, we
start from an initial state $|\psi_k^{\rm in}\rangle$ and compute the
expectation value
\begin{eqnarray}
C_k(nT) &=& \langle \psi_k^{\rm in}|(U^{\dagger}_k)^n O_k (U_k)^n
|\psi_k^{\rm in}\rangle, \label{corrfn1} \end{eqnarray}
where $O_k$ is a generic quadratic operator constructed out of
$\psi_k$ and $\psi_k^{\dagger}$. The specific forms of these
operators shall be discussed in subsequent sections in the context
of the SSH and Ising models. We note that for the integrable models
treated here, the correlations of $O_k$ constitute the most general
independent correlation functions; all quartic or higher order
correlation of fermionic operators can be expressed in terms of $O_k$.

Using Eq.\ \eqref{evolop}, we can express these correlations as
\begin{eqnarray}
C_k(nT) &=& C_{0k} + \delta C_k(nT), \non \\
C_{0k} &=& \sum_j |\alpha_j(k)|^2 O_{jj}(k), \nonumber\\
\delta C_k(nT) &=& e^{-i n \Delta (k) T/\hbar} f(k) + {\rm H.c.}, \nonumber\\
f(k) &=& \alpha_2^{\ast}(k) \alpha_1(k) O_{12}(k), \nonumber\\
C_x(nT) &=& \int_{\rm BZ} \frac{dk}{2\pi} e^{i k x} C_k(nT), \label{corrdef}
\end{eqnarray}
where the integral is taken over the Brillouin zone. Here the
Floquet energy gap $\Delta(k)$, the overlap $\alpha_{j}(k)$ of the
initial state with the Floquet eigenstates, and the matrix elements
$O_{j_1 j_2}(k)$ are given by
\begin{eqnarray}
\Delta(k) &=& \epsilon_F^{(1)}(k)-\epsilon_F^{(2)}(k), \quad
\alpha_j(k)= \langle \psi_k^{\rm in}|n_j(k)\rangle, \nonumber\\
O_{j_1 j_2}(k) &=& \langle n_{j_1}(k)|O_k|n_{j2}(k)\rangle.
\label{deldef}
\end{eqnarray}
We note that the Fourier transform of $C_{0k}$ (Eq.\ \eqref{corrdef})
denotes the steady state value of $C_x$ in real space which is
independent of $n$. Thus
\begin{eqnarray}
\delta C_x(nT) = \int_{\rm BZ} \frac{dk}{2\pi} e^{i k x} (f(k)e^{-i
n \Delta(k) T/\hbar} +{\rm H.c.}) \label{corrdev} \end{eqnarray}
represents the deviation of $C_x (nT)$ from its steady state value
in real space. Since such a steady state is reached for large $n$ in
any driven system, we expect $\delta C_x(nT)$ to be a decaying
function of $n$ for large $n$.

To understand the nature of this decay, we note that for large $n$,
the integral for $\delta C_x(nT)$ can be evaluated within a
stationary point approximation. To this end, let us assume that the
leading contribution to the integral comes from a stationary point
at $k=k_0$. Around this point, let us assume that
\begin{eqnarray}
\Delta (k) &\simeq& \Delta(k_0) + \Delta^{(\beta)}(k_0) \delta k^{\beta}
+ \cdots \nonumber\\
f(k) &\simeq& f(k_0) + f^{\alpha}(k) \delta k^{\alpha} + \cdots
\nonumber\\
\Delta^{(\beta)}(k_0) &=& \frac{\partial^{\beta}\Delta(k)}{\partial
k^{\beta}}\Big|_{k=k_0}, \quad f^{\alpha}(k_0) =
\frac{\partial^{\alpha} f(k)}{\partial k^{\alpha}}\Big|_{k=k_0}, \non \\
&& \label{expans} \end{eqnarray} where $\alpha$ and $\beta$ denote
the leading powers for expansion $\Delta(k)$ and $f(k)$ respectively
around $k=k_0$. We note that since $k_0$ is a stationary point,
$\beta \ge 2$. Substituting Eq.\ \eqref{expans} in Eq.\
\eqref{corrdev}, we find the leading behavior of the correlation to
be given by
\begin{eqnarray}
\delta C_x(nT) &\sim& A(k_0;n,T) + \int_{-\infty}^{\infty} \frac{d
\delta k}{2\pi} e^{- i \delta k x} \nonumber\\
&& \times (f^{\alpha}(k_0) \delta k^{\alpha} e^{-i n
\Delta^{(\beta)}(k_0) \delta k^{\beta} T/\hbar} + {\rm H.c.}),\non \\
&& \label{deltaeq} \end{eqnarray} where we have included $f(k_0)
\equiv f^{(0)}(k_0)$ by allowing the exponent $\alpha$ to have zero
value in the second term. Here $A(k_0;n,T)$ is the value of the
integral obtained from the first term of the stationary point
expansion. This term is non-zero if $f(k_0)$ is finite, but it does
not contribute to the decay of the correlator since its an
oscillatory function of $n$. A scaling $ \delta k \to \delta k'=
n^{1/\beta} \delta k$ and $x \to x'= x/n^{1/\beta}$ in the integral
in Eq.\ \eqref{deltaeq} leads to
\begin{eqnarray}
\delta C_x(nT) &=& A(k_0;n,T) + n^{-(\alpha+1)/\beta} g(k_0;x'),
\label{corrdecay} \\
g(k_0;z) &=& \int_{-\infty}^{\infty} \frac{dy}{2 \pi} f^{\alpha}(k_0)
 y^{\alpha} e^{i(y z- \Delta^{(\beta)}(k_0) y^{\beta} T/\hbar)} + {\rm H.c.}
\nonumber \end{eqnarray}
Since $g(k_0;z)$ is an oscillatory function of $z$, it does not
contribute to the decay of the correlators. Thus we find the general
result that the leading decay of the correlator is given by
\begin{eqnarray} C_x(nT) \sim n^{-(\alpha+1)/\beta} \label{final}
\end{eqnarray}
which is the main result of this section. For multiple stationary
points, it is easy to see that the leading behavior is given by the
one which allows for the slowest decay.

We note that for any stationary point expansion, generically, we
expect $\beta=2$ since the second derivative of the energy gap need
not vanish at the stationary point. In this case, we find that the
correlators would decay as $\delta C_x(nT) \sim n^{-3/2}$ if
$f(k_0)$ vanishes at the  point and $f(k_0+\delta k) \sim (\delta
k)^2$ as $\delta k \rightarrow 0$ and as $\delta C_x(nT) \sim
n^{-1/2}$ if $f(k_0)$ is finite (this corresponds to $\alpha=0$).
These two behaviors correspond to two dynamical phases; the former
behavior is seen for high drive frequencies where the stationary
point typically occurs at the edge of the Brillouin zone
\cite{asen1}, while the latter occurs at lower frequencies where
additional stationary points which correspond to $\alpha=0$ appear
inside the Brillouin zone. As noted in Ref.\ \onlinecite{asen1},
these two phases are separated by a dynamical phase transition
characterized by a critical drive frequency $\omega_c$.

The decay of the correlators exactly at the transition allows for
richer behavior which we explore next. We note that exactly at the
transition point, the Floquet energy gap $\Delta(k)$ must have a
point of inflection which necessitates its second derivative to also
vanish. Thus for this case $\beta>2$. Depending on the symmetry of
model, we find that either the third or the fourth derivative of the
Floquet gap contributes to the lowest non-vanishing term in the
expansion of $\Delta(k)$ about $k=k_0$. The former behavior
corresponds to $\beta=3$ and occurs if the Floquet energy is odd
under the transformation $k \to -k$. This leads to
\begin{eqnarray} C_x(nT) \sim n^{-(\alpha+1)/3}. \label{odddecay}
\end{eqnarray}
In contrast, if the Floquet energy is even under $k \to -k$, its
fourth derivative contributes to the lowest non-vanishing term. This
yields $\beta=4$ and leads to
\begin{eqnarray} C_{x}(nT) \sim n^{-(\alpha+1)/4} \label{evendecay}.
\end{eqnarray}
Thus the decay of the correlators may follow a different power law
at the critical point between two dynamical phases. We note that the
presence of two distinct dynamical phases across a transition is a
sufficient condition for such behavior; however it is not a
necessary condition and we shall discuss an example of such
anomalous decay without the presence of distinct dynamical phases in
the next section.

Finally, we discuss the crossover scale $n_c$ which
denotes the number of drive cycles after which the system crosses
over to a decay characterized by $\beta=2$. We note that $n_c$
diverges at a dynamical transition and tends to zero far away
from it. To estimate $n_c$, we note that near a transition we can
always write
\begin{eqnarray}
&& \delta C_x(nT) \sim A(k_0;n,T) + \int_{-\infty}^{\infty} \frac{d
\delta
k}{2\pi} e^{- i \delta k x} \label{cross1} \\
&& \times (f^{\alpha}(k_0) \delta k^{\alpha} e^{-i n ( c_1 \delta
k^{a_0} + c_2 \delta k^{\beta_0}+...) T/\hbar} + {\rm H.c.}), \nonumber
\end{eqnarray}
where $c_1$ and $c_2$ are the coefficients of expansions of the
Floquet spectrum around $k=k_0$, $\beta_0$ denotes the lowest
integer larger than $a_0$ for which $c_2 \ne 0$, $a_0=1$ or $2$
depending on the symmetry of the Floquet spectrum, and the ellipsis
indicates higher order terms in the expansion of $\Delta(k)$ around $k=k_0$
which we shall ignore. A simple scaling $ k \to \delta k' =
n^{1/\beta_0} \delta k$ yields
\begin{widetext}
\begin{eqnarray}
\delta C_x(nT) &\sim& A(k_0;n,T) + \int_{-\infty}^{\infty} \frac{d
\delta k'}{2\pi} e^{- i \delta k' n^{1/\beta_0} x} [f^{\alpha}(k_0)
n^{-(\alpha+1)/\beta_0} (\delta k')^{\alpha} e^{-i ( c_1
n^{1-a_0/\beta_0} (\delta k')^{a_0} + c_2 (\delta k')^{\beta_0}) T/\hbar}
\nonumber\\
&& ~~~~~~~~~~~~~~~~~~~~~~~~~~~~~~~~~~~~~~~~~~~~~~+ {\rm H.c.}].
\label{cross2} \end{eqnarray}
\end{widetext}
Thus the behavior of the integral is governed by the coefficient of
$(\delta k')^{a_0}$ in the exponent after
\begin{eqnarray} n_c \simeq (c_2/c_1)^{\beta_0/(\beta_0-a_0)} \label{cros3}
\end{eqnarray}
drive cycles. Hence the crossover scale is also controlled by the
symmetry of the model which renders $a_0=1(2)$ and $\beta_0=3(4)$
for models whose Floquet spectrum is odd (even) under $k \to -k$
near the transition point. Furthermore, for a generic transition
point for these integrable models $c_1 \sim |\omega - \omega_c|$ and
$c_2$ is a constant. Thus we find
\begin{eqnarray}
n_c \sim |\omega-\omega_c|^{-\beta_0/(\beta_0-a_0)} \label{cros4}
\end{eqnarray}
which shows that $n_c$ diverges at the transition point where
$c_1=0$ and it is small away from the transition where generically
$c_1 \ll c_2$. We explore this crossover physics in detail in Secs.\
\ref{ssh} and \ref{ising} in the context of specific models.

\section{SSH model}
\label{ssh}

In this section, we will study the effect of periodic driving in the
Su-Schrieffer-Heeger (SSH) model. We will show that the long-time
behavior of the correlation function can show transitions between
different power laws for some special choices of the driving
parameters. We shall analyze the driven SSH model within first-order
FPT; this is done so as to obtain simple analytical insights. The
results obtained from FPT shall be compared with exact numerics
towards the end of the section.

The SSH model is a tight-binding model of non-interacting electrons
in 1D in which the nearest-neighbor hopping has different strengths
on alternate bonds \cite{sshref}. We will ignore the spin of the
electron since it will not play any role in this paper. In
second-quantized notation, the Hamiltonian for a system with $N$
sites (where $N$ is even) and periodic boundary conditions is given
by \beq H ~=~ \sum_{n=1}^{N/2} ~[\ga_1 a_n^\dagger b_n ~+~ \ga_2
b_n^\dagger a_{n+1} ~+~ {\rm H.c.}], \label{hamssh1} \eeq where
$a_{N/2+1} \equiv a_1$. (We will set both Planck's constant $\hbar$
and the spacing $a$ between nearest-neighbor sites to 1).
Transforming to momentum space, we find that \beq H ~=~ \sum_k
~[\ga_1 a_k^\dagger b_k ~+~ \ga_2 b_k^\dagger a_k e^{i2k} ~+~ {\rm
H.c.}], \label{hamssh2} \eeq where $k$ takes $N/2$ equally spaced
values lying in the range $[-\pi/2,\pi/2]$. This can be written in
terms of a $2 \times 2$ matrix $H_k$ as \bea H &=& \sum_k ~\left(
\begin{array}{cc} a_k^\dagger & b_k^\dagger \end{array} \right)
~H_k~ \left( \begin{array}{c}
a_k \\
b_k \end{array} \right), \non \\
H_k &=& \left( \begin{array}{cc}
0 & \ga_1 + \ga_2 e^{-i2k} \\
\ga_1 + \ga_2 e^{i2k} & 0 \end{array} \right). \label{hamssh3} \eea

The energy-momentum dispersion is given by $E_{k\pm} = \pm E_k$,
where \beq E_k ~=~ \sqrt{\ga_1^2 ~+~ \ga_2^2 ~+~ 2 \ga_1 \ga_2 \cos
(2k)}. \label{disp1} \eeq We see that the spectrum has a minimum gap
equal to $E_{k+} - E_{k-} = 2 |\ga_1 \pm \ga_2|$ at $k = 0$ and $\pm
\pi/2$ respectively, depending on whether $\ga_1$ and $\ga_2$ have
opposite signs or the same sign.

We will now consider driving this system periodically
in time~\cite{asboth,zheng,bala1,bala2,yates,borja} by adding a
term to the hopping which is of the form $a \sin (\om t)$, where $a$ and
$\om$ are the driving amplitude and frequency respectively.
The Hamiltonian in momentum space is therefore given by
\bea H &=& \sum_k ~[(\ga_1 + a \sin (\om t)) a_k^\dagger b_k \non \\
&& ~~~~~~+~ (\ga_2 + a \sin (\om t) b_k^\dagger a_k e^{i2k} ~+~ {\rm
H.c.}]. \label{hamssh4} \eea

This system can be analytically studied by several methods such as
the Floquet-Magnus expansion which works in the limit $\om$ is much
larger than all the other parameters, $a$, $\ga_1$ and $\ga_2$, and
FPT which is valid in the limit that
both $a$ and $\om$ are much larger than $\ga_1$ and $\ga_2$. We will
use FPT which proceeds as follows.

For each value of $k$, we consider the Floquet operator \beq U_k ~=~
{\cal T} ~\exp [-i \int_0^T dt H_k (t)]. \label{uk} \eeq where $\cal
T$ denotes time-ordering. Note that $U_k$ is an $SU(2)$ matrix since
$H_k (t)$ is a Hermitian and traceless matrix for all times $t$. We
can write the Floquet operator as \beq U_k ~=~ e^{-i H_{Fk} T},
\label{hf1} \eeq where $H_{Fk}$ is time-independent and is called
the Floquet Hamiltonian. Assuming $a \gg \ga_1, ~\ga_2$, we write
\bea H_k (t) &=& H_0 (t) ~+~ V, \non \\
H_0 (t) &=& \left( \begin{array}{cc}
0 & a \sin (\om t) (1 + e^{-i2k}) \\
a \sin (\om t) (1 + e^{i2k}) & 0 \end{array} \right), \non \\
V &=& \left( \begin{array}{cc}
0 & \ga_1 + \ga_2 e^{-i2k} \\
\ga_1 + \ga_2 e^{i2k} & 0 \end{array} \right). \label{hamssh5} \eea
We will find the form of $H_{Fk}$ only to first order in the perturbation $V$.

The instantaneous eigenvalues of $H_0 (t)$ are given by $E_{k+}
=2a\sin (\om t) \cos k$ and $E_{k-} =-2a\sin (\om t) \cos k$. These
satisfy the condition \bea e^{i \int_{0}^{T}dt\left( E_{k+} - E_{k-}
\right)}=1. \eea We will therefore have to carry out degenerate FPT.
The eigenfunctions corresponding to $E_{k\pm}$ are given by \bea
\ket{+}_k &=& \frac{1}{\sqrt{2}} \left(\begin{array}{cc}
1 \\ e^{ik} \end{array}\right), \non \\
\ket{-}_k &=& \frac{1}{\sqrt{2}} \left(\begin{array}{cc} 1 \\
-e^{ik} \end{array}\right). \eea

We begin with the Schr\"odinger equation \beq
i\frac{d\ket{\psi}}{dt} ~=~ \left( H_{0}+V \right)\ket{\psi}.
\label{schr1} \eeq We assume that $\ket{\psi (t)}$ has the expansion
\bea \ket{\psi(t)} ~=~ \sum_{n} ~c_{n}(t)
~e^{-i\int_{0}^{t}dt'E_{n}(t')} \ket{n}, \label{psi1} \eea where
$\ket{n}=\ket{+}$ and $\ket{-}$. Eq.~\eqref{schr1} then implies that
\beq \frac{dc_{m}}{dt} ~=~ -i ~\sum_{n} ~\bra{m} V\ket{n}
~e^{i\int_{0}^{t} dt' \left( E_{m}(t')-E_{n}(t')\right)} c_{n}.
\label{cm1} \eeq

Integrating this equation, and keeping terms only to first order in
$V$, we find that \bea c_{m}(T) &=& c_{m}(0) ~- ~i
~\sum_{n}\int_{0}^{T}dt\bra{m} V\ket{n} \non \\
&& \times ~e^{i\int_{0}^{t} \left(E_{m}(t')-
E_{n}(t')\right)~dt'}c_{n}(0). \label{cm2} \eea This can be written
as \bea c_{m}(T) ~=~
\sum_{n}\left(I-iH_{Fk}^{(1)}T\right)_{mn}c_{n}(0), \label{hf2}
\label{cm3} \eea where $I$ denotes the identity matrix and
$H_F^{(1)}$ is the Floquet Hamiltonian to first order in $V$. We
then find that \bea \bra{+} H_{Fk}^{(1)} \ket{+}&=&\left(
\ga_{1}+\ga_{2} \right)
\cos k,\non\\
\bra{-}H_{Fk}^{(1)} \ket{-}&=&-\left( \ga_{1}+\ga_{2} \right)
\cos k,\non\\
\bra{+}H_{Fk}^{(1)} \ket{-}&=&-i\left( \ga_{1}-\ga_{2} \right)
\sin k J_{0}\left(\frac{4a}{\omega}\cos k\right),\non\\
\bra{-}H_{Fk}^{(1)} \ket{+}&=&i\left( \ga_{1}-\ga_{2} \right)
\sin k J_{0}\left(\frac{4a}{\omega}\cos k\right). \eea
$H_{Fk}^{(1)}$ then takes the following form in the
$\ket{+}_k,~\ket{-}_k$ basis
\bea H_{Fk}^{(1)} &=& (\ga_{1}+\ga_{2}) \cos k ~\si_z \non \\
&& +~ (\ga_{1}-\ga_{2}) \sin k
J_{0}\left(\frac{4a}{\omega}\cos k\right) \si_y. \eea

We now change basis to
\bea \ket{\uparrow}_{k}&=&a_{k}^{\dagger}\ket{0},\non\\
\ket{\downarrow}_{k}&=&b_{k}^{\dagger}\ket{0}, \eea so that \bea
\ket{+}_k &=& \frac{1}{\sqrt{2}}\left( \ket{\uparrow}_{k}+e^{ik}
\ket{\downarrow}_{k}\right),\non\\
\ket{-}_k &=& \frac{1}{\sqrt{2}}\left( \ket{\uparrow}_{k}-e^{ik}
\ket{\downarrow}_{k}\right). \eea
\begin{widetext}
In the $\ket{\uparrow}_{k}$, $\ket{\uparrow}_{k}$ basis, we get \bea
H_{Fk}^{(1)}&=&\left[\left(\ga_{1}+\ga_{2}\right)\cos^{2}k+
\left(\ga_{1}-\ga_{2}\right)J_{0}\left(\frac{4a}{\omega}\cos
k\right)
\sin^{2}k\right]\si_{x}\non\\
&&+\sin k \cos
k\left[\left(\ga_{1}+\ga_{2}\right)-\left(\ga_{1}-
\ga_{2}\right)J_{0}\left(\frac{4a}{\omega}\cos k\right)\right]
\si_{y}. \label{hf3} \eea
\end{widetext}

Before proceeding further, we make two comments about the exact form
of $H_{Fk}$ to all orders based on certain symmetries. First,
$H_{Fk}$ must be an odd function of $\ga_1, ~\ga_2$. To see this, we
note that Eq.~\eqref{uk} can be written as a product of $N_t$
factors in which $t$ increases from 0 to $T$ as we go from right to
left in steps of $T/N_t$ (eventually we take the limit $N_t \to
\infty$). We then use the fact that the driving term satisfies $\sin
(\om (T-t)) = - \sin (\om t)$ to see that \beq [U_k (\ga_1,
\ga_2)]^{-1} ~=~ U_k (-\ga_1,-\ga_2), \label{sym1} \eeq if we hold
$a, ~\om$ fixed. Eq.~\eqref{hf1} then implies that \beq H_F (-\ga_1,
-\ga_2) ~=~ - H_F (\ga_1, \ga_2). \label{sym2} \eeq Hence $H_F$ can
only have odd powers of $\ga_1, ~\ga_2$. This implies that if
$\ga_1, \ga_2 \ll a, \om$, the first-order Floquet Hamiltonian will
be a very good approximation to the exact Floquet Hamiltonian since
the next correction is of third order in $\ga_1, ~\ga_2$. Second,
let us consider the special case $\ga_2 = - \ga_1$ which will be
considered in more detail below. We then find that after doing a
unitary transformation,
\bea H_k (t) &\to& V_k H_k (t) V_k^\dagger, \non \\
{\rm where} ~~~V_k &=& \left( \begin{array}{cc}
1 & 0 \\
0 & e^{-ik} \end{array} \right), \eea we obtain \beq H_k (t) ~=~ 2 a
\sin (\om t) \cos k ~\si_x ~-~ 2 \ga_1 \sin k ~ \si_y.
\label{hamssh6} \eeq We now use Eq.~\eqref{hamssh6} to calculate
the Floquet operator in and Floquet Hamiltonian in Eqs.~\eqref{uk}
and ~\eqref{hf1}. Then an argument similar to the one above shows
that \beq [U_k]^{-1} ~=~ U_{-k}, \label{sym3} \eeq where we have
held $\ga_1, ~\ga_2$ fixed and only changed $k \to - k$.
Eq.~\eqref{sym3} implies that \beq H_{F,-k} ~=~ - H_{F,k}.
\label{sym4} \eeq This means that the eigenvalues of $H_F$
(quasienergies) must be odd functions of $k$ if $\ga_2 = - \ga_1$.

\begin{figure}[h!]
\begin{center}
\includegraphics[height=5.9cm]{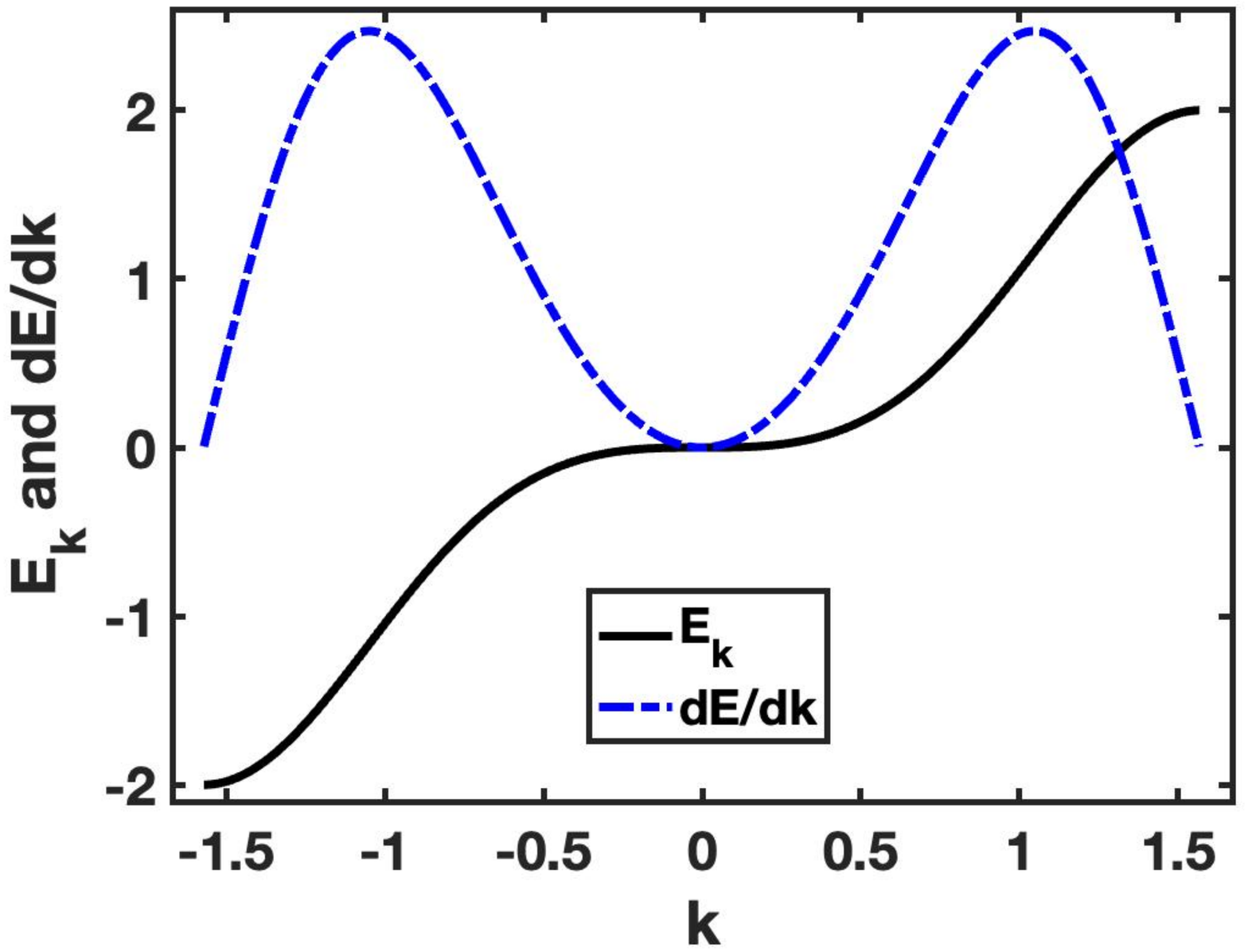}
\end{center}
\caption{Energy $E_k$ (black solid line) and its first derivative
$dE_k/dk$ (blue dot-dashed line) versus $k$ obtained from the
first-order Floquet Hamiltonian for $\ga_1 = 1, ~\ga_2 = -1, ~a = 6$
and $\om = 4 a /\mu_1$. There is a stationary point at $k=0$ with
both $dE_k/dk$ and $d^2E_k/dk^2$ equal to zero.} \label{sshfig01}
\end{figure}

The eigenvalues of the first-order Floquet Hamiltonian in
Eq.~\eqref{hf3} are given by $\pm E_k$ (with the Floquet energy gap
being $\Delta(k)=2 E(k)$) where
\begin{widetext}
\beq E_k ~=~ \sqrt{(\ga_1 ~+~ \ga_2)^2 \cos^2 k ~+~ (\ga_1 ~-~
\ga_2)^2 \sin^2 k \left[ J_{0}\left(\frac{4a}{\omega}\cos k\right)
\right]^2}. \eeq
\end{widetext}
For general values of $\ga_1, ~\ga_2$, we see that $E_k$ is non-zero
for all values of $k$. However, for $\ga_2 = \ga_1$ it vanishes if
$k = \pi/2$ (in fact, $E_k$ does not depend on the driving if $\ga_2
= \ga_1$), while for $\ga_2 = - \ga_1$ it vanishes when either $k=0$
or $J_0 ((4 a /\om) \cos k) = 0$. Thus driving can lead to
non-trivial zeros of the Floquet energy for special values of
$(a/\om) \cos k$. In the rest of this section we will therefore
consider the case $\ga_2 = - \ga_1$. Setting $\ga_1 = 1$, we have
\beq E_k ~=~ 2\sin k ~J_{0}\left(\frac{4a}{\omega}\cos k\right).
\label{ek1} \eeq Fig.~\ref{sshfig01} shows plots of $E_k$ and
$dE_k/dk$ for a system with $\ga_1 = 1, ~\ga_2 = -1, ~a = 6$ and
$\om = 4 a /\mu_1$, where $\mu_1 ~\simeq~ 2.4048$ is the first zero
of $J_0 (z)$.

We now consider an operator of the form $a_j^\dagger b_j$ where $j$
denotes a particular unit cell. Starting from an initial state $\Psi
(0)$, we will look at the correlation function at stroboscopic
instances of time $t=nT$, \beq C_n ~=~ \bra{\Psi (nT)} a_j^\dagger
b_j \ket{\Psi (nT)}, \label{cn1} \eeq and we will study how this
behaves for large values of $n$. We take the initial state to be a
half-filled state given by a product in momentum space \beq \Psi (0)
~=~ \prod_{k} ~[(a_k^\dagger ~+~ e^{i \phi} b_k^\dagger) / \sqrt{2}]
~\ket{{\rm vac}}. \label{psiinit} \eeq For simplicity we have taken
the phase $\phi$ to be independent of $k$.

\bea C_n&=&\frac{2}{N} ~\sum_{k} ~\bra{\Psi(nT)}a_{k}^{\dagger}b_{k}
\ket{\Psi(nT)}\non\\
&=&\frac{2}{N} ~\sum_{k}~
\bra{\Psi(0)} (U_{k}^{\dagger})^n a_{k}^{\dagger}b_{k}
(U_{k})^n \ket{\Psi(0)}.\non\\ \eea $H_{Fk}^{(1)}$ can be written in
the following matrix form \bea
H_{Fk}^{(1)}&=&E_{k}\left(\begin{array}{cc} 0 & ie^{-ik}\\ -ie^{ik}
& 0\end{array}\right), \label{hf4} \eea whose eigenvalues are $\pm
E_{k}$ and eigenfunctions are \bea
\ket{+}_{k}&=&\frac{1}{\sqrt{2}}\left(\begin{array}{cc}
1 \\ -ie^{ik}\end{array}\right),\non\\
\ket{-}_{k}&=&\frac{1}{\sqrt{2}}\left(\begin{array}{cc} 1 \\
ie^{ik}\end{array}\right). \eea Then we have \bea (U_{k})^n &=&
e^{-inE_{k}T}\ket{+}_{k}\bra{+}_{k}+e^{inE_{k}T}\ket{-}_{k}
\bra{-}_{k}.\non\\
&=& \left(\begin{array}{cc} \cos (nE_{k}T) & e^{-ik}\sin (nE_{k}T)\\
-e^{ik}\sin (nE_{k}T) & \cos (nE_{k}T)\end{array}\right).\non\\\eea
Since \bea a_{k}^{\dagger}=\left(\begin{array}{cc} 1 \\
0\end{array}\right), ~~~b_{k}^{\dagger}=\left(\begin{array}{cc} 0\\
1\end{array}\right),\eea we find that
\begin{widetext}
\bea (U_{k}^{\dagger})^n a_{k}^{\dagger}b_{k} (U_{k})^n ~=~
\left(\begin{array}{cc}
-\frac{1}{2} e^{ik} \sin (2nE_{k}T) & \cos^{2}(nE_{k}T)\\
-e^{i2k}\sin^{2}(nE_{k}T) & \frac{1}{2} e^{ik}
\sin (2nE_{k}T)\end{array} \right). \label{uknt} \eea
\end{widetext}
Using Eqs.~\eqref{psiinit} and \eqref{uknt}, we obtain
\bea C_{n}&=& A ~+~ \frac{2}{N} ~\sum_{k} ~f(k) \cos (2nTE_{k}),\non\\
A&=& \frac{1}{2N} ~\sum_{k}
~\left(e^{i\phi}-e^{i\left(2k-\phi\right)}\right),
\non\\
f(k)
&=&\frac{1}{4}~\left(e^{i\phi}+e^{i\left(2k-\phi\right)}\right).
\eea For $N \to \infty$, these quantities have the integral forms
\bea C_{n}&=& A ~+~ \frac{1}{4\pi} ~\int_{-\pi/2}^{\pi/2} dk ~\left(
e^{i \phi} + e^{-i \phi} \cos (2k) \right) \non \\
&& ~~~~~~~~~~~~~~~~~~~~~~~~~~~\times \cos (2nTE_{k}), \non\\
A&=&\frac{1}{4\pi} ~\int_{-\pi/2}^{\pi/2} dk ~\left(e^{i\phi}-
e^{i\left(2k-\phi \right)}\right) ~=~ \frac{e^{i \phi}}{4},
\label{caf} \eea where we have used the relation
$\cos (2nTE_{k})=\cos (2nTE_{-k})$ (since $E_{-k} = - E_k$) to write
the first equation in Eq.~\eqref{caf}.

We will now study the form of the $n$-dependent part of $C_n$, called
$\delta C_n$, for large $n$. The dominant
contributions will come from regions around the values of $k$ where
$E_k$ has an extremum, namely, $dE_k /dk = 0$. One such point is $k=
\pi/2$. Expanding around it to second order, we find that $E_k = 2 -
(1 + 8a^2 /\om^2) (k - \pi/2)^2$, where we have used the expansion
$J_0 (z) = 1 - z^2/4$ for small $z$. We first assume that $f(k =
\pi/2) \ne 0$; this will be true if $\phi$ is not an integer
multiple of $\pi$. Near $k' = k - \pi/2$, the $n$-dependent term in
Eq.~\eqref{caf} then takes the form
\bea \delta C_n &\simeq& \frac{i}{2 \pi} ~\int dk' ~\sin \phi \non \\
&& \times ~{\rm Re} ~\exp [i 4 n T - i 2n T (1 + 8a^2 / \om^2)
k'^2], \label{cn2} \eea where Re denotes real part. We thus see that
$C_n$ will oscillate as $\cos (4n T)$ (which implies that its
absolute value will vary periodically with $n$ with a period $\Delta
n = \pi/(4 T) = \om /8$) multiplied by an integral of the form $\int
dk' \exp [i \al n k'^2]$ which, by a scaling argument, will decay as
$1/n^{1/2}$ for large $n$. However, in the special case that $\phi$
is an integer multiple of $\pi$, both $e^{i \phi} + e^{-i \phi} \cos
(2k)$ and its first derivative vanish at $k=\pi/2$. We then get a
factor of $k'^2$ appearing in the integrand of Eq.~\eqref{cn2}. The
integral will therefore be of the form $\int dk' k'^2 \exp [i \al n
k'^2]$ which will decay as $1/n^{3/2}$ for large $n$. Below we will
see plots showing a $1/n^{1/2}$ decay (for $\phi = \pi/4$) and
a $1/n^{3/2}$ decay (for $\phi = 0$).

Next, we consider if there are any other values of $k$ where $dE_k
/dk = 0$. We find that such points exist if $\om < \om_1$, where
$\om_1 = 4 a /\mu_1$ with $\mu_1 \simeq 2.4048$ being the first zero
of $J_0 (z)$. This is because, as $k$ goes from 0 to $\pi/2$, $\sin
k$ goes from 0 to 1, taking only positive values in between, while
$J_0 ((4 a/\om) \cos k)$ goes between $J_0 (4 a/\om)$ and 1,
crossing zero $p$ times in between if $4 a /\om$ is larger than the
first $p$ zeros of $J_0 (z)$. This implies that $E_k$ in
Eq.~\eqref{ek1} will go between 0 and 2, crossing zero $p$ times in
between; hence $E_k$ will have $p$ extrema where $dE_k /dk = 0$.
Next, if $k=k_0$ is one of the points where $dE_k /dk = 0$, and
$k_0$ is not equal to either 0 or $\pi/2$, the factor $f(k)$ in
Eq.~\eqref{cn1} is not zero, but the argument of $\cos (2 E_k n T)$
will go as $\cos (2 E_{k_0} n T + \al n (k - k_0)^2)$. This means
that $\delta C_n$ will oscillate as $\cos (2 E_{k_0} n T)$ (implying that
its absolute value with vary periodically with a period $\Delta n =
\om /(4 E_{k_0})$) multiplied by an integral of the form $\int dk'
\exp [i \al n k'^2]$. By a scaling argument, this will again decay
as $1/n^{1/2}$ for large $n$.

\begin{widetext}
\begin{figure*}[!tbp]
\includegraphics[width=0.335\hsize]{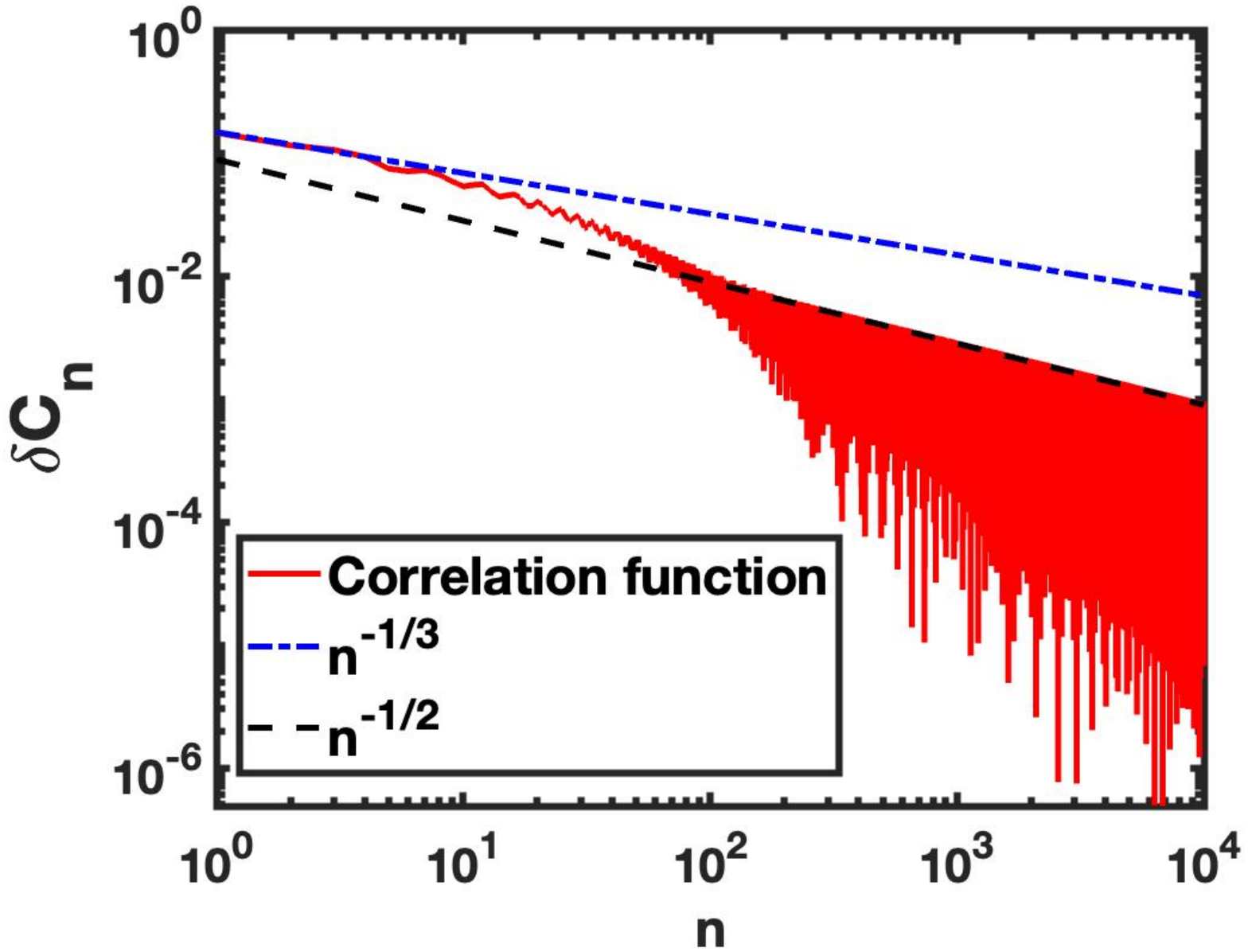}%
\includegraphics[width=0.335\hsize]{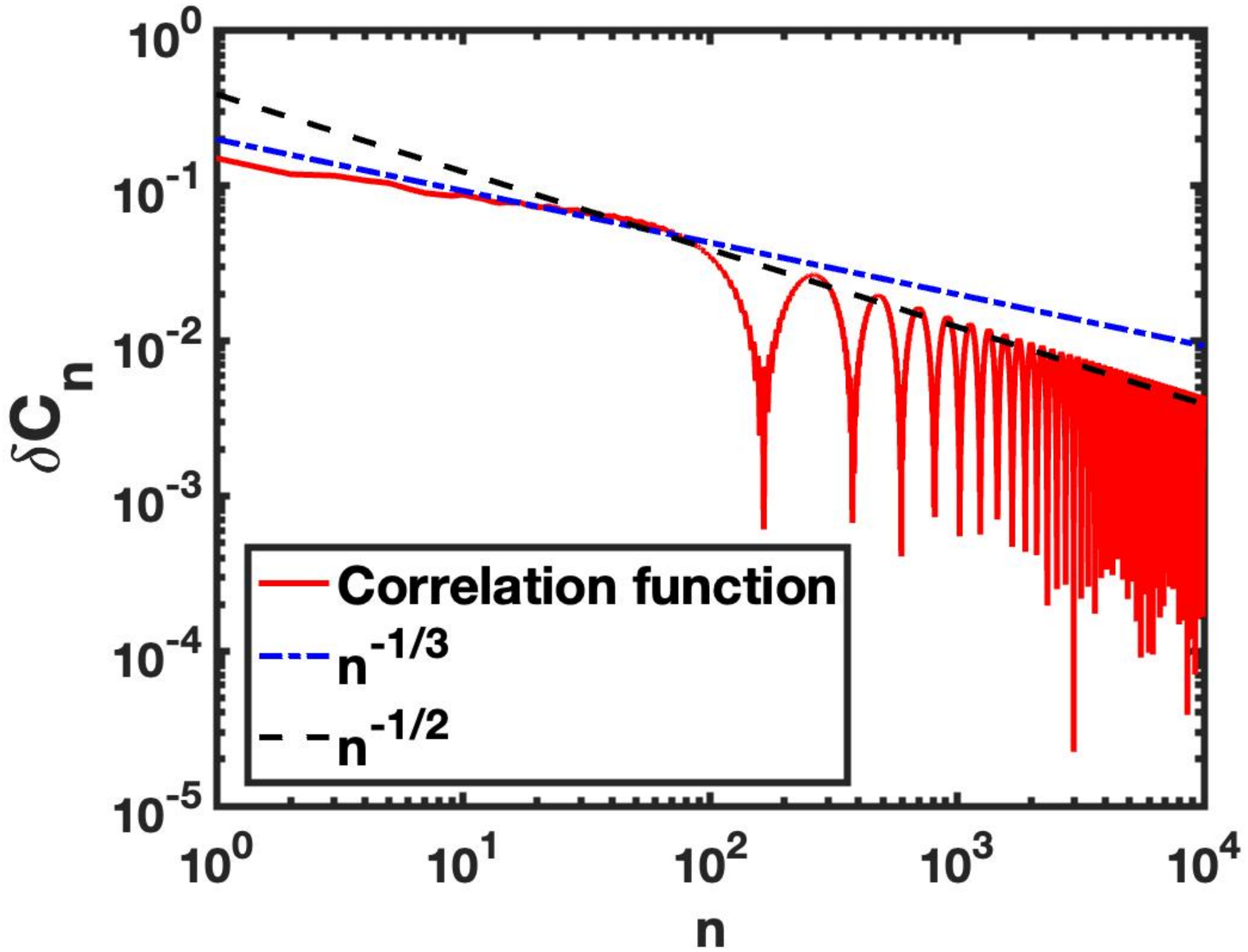}%
\includegraphics[width=0.335\hsize]{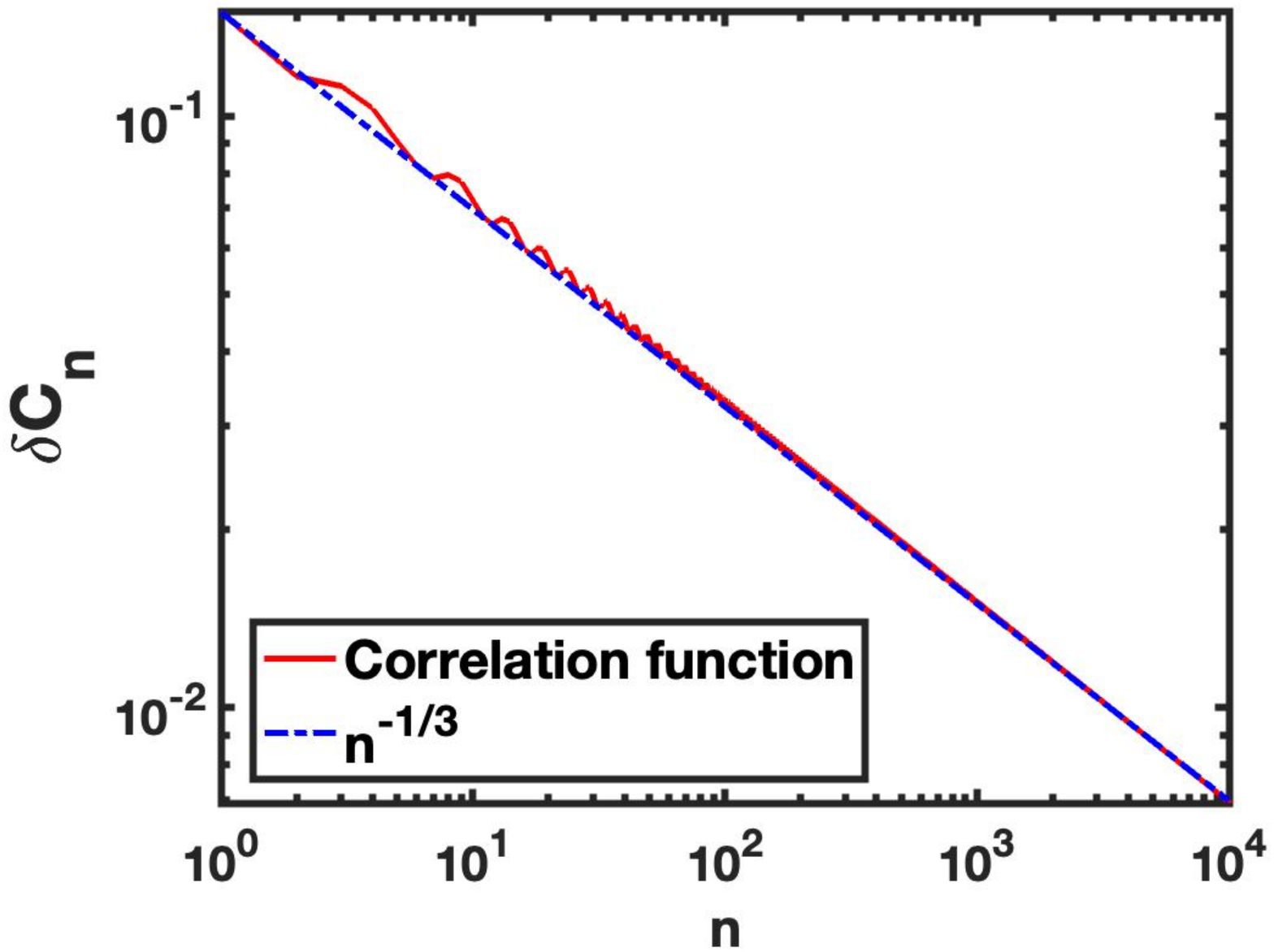}%
\caption{Log-log plots of the absolute value of the $n$-dependent part
$\de C_n$ of the correlation function
$\langle a_j^\dagger b_j \rangle$ as a function of the time $nT$, for
$\ga_1 = 1, ~\ga_2 = -1, ~a = 6$, $\om = 4 a /(\mu_1 + \ep)$, and
$\phi = \pi/4$ for the initial state. Left panel: $\ep = -0.1$, so that $\om >
\om_1$. Middle panel: $\ep = 0.1$, hence $\om < \om_1$. Right panel:
$\ep = 0$, hence $\om = \om_1$. Both the left and the middle panels
show crossovers between $1/n^{1/3}$ and an oscillating function
times $1/n^{1/2}$. The right panel with $\ep = 0$ and hence $\om =
\om_1$ shows only a $1/n^{1/3}$ scaling with no oscillations.} \label{sshfig02}
\end{figure*}
\end{widetext}

We thus conclude that $\delta C_n$ will generally oscillate and decay as
$1/n^{1/2}$.
This is what we see in Figs.~\ref{sshfig02} (a) and (b) for a system with
$\ga_1 = 1, ~\ga_2 = -1, a = 6$, $\om = 4 a /(\mu_1 \pm \ep)$, where $\ep =
0.1$, and $\phi = \pi/4$ for the initial state (see Eq.~\eqref{psiinit}).
If $4 a/\om$ is larger than
$p$ zeros of $J_0 (z)$ (where $p$ can be $1, 2, 3, \cdots$), there
will be $p$ terms in $\delta C_n$ all of which decay as $1/n^{1/2}$ but
which oscillate with $p$ different periods $\Delta n$.

\begin{figure}[h!]
\begin{center}
\includegraphics[height=6.2cm]{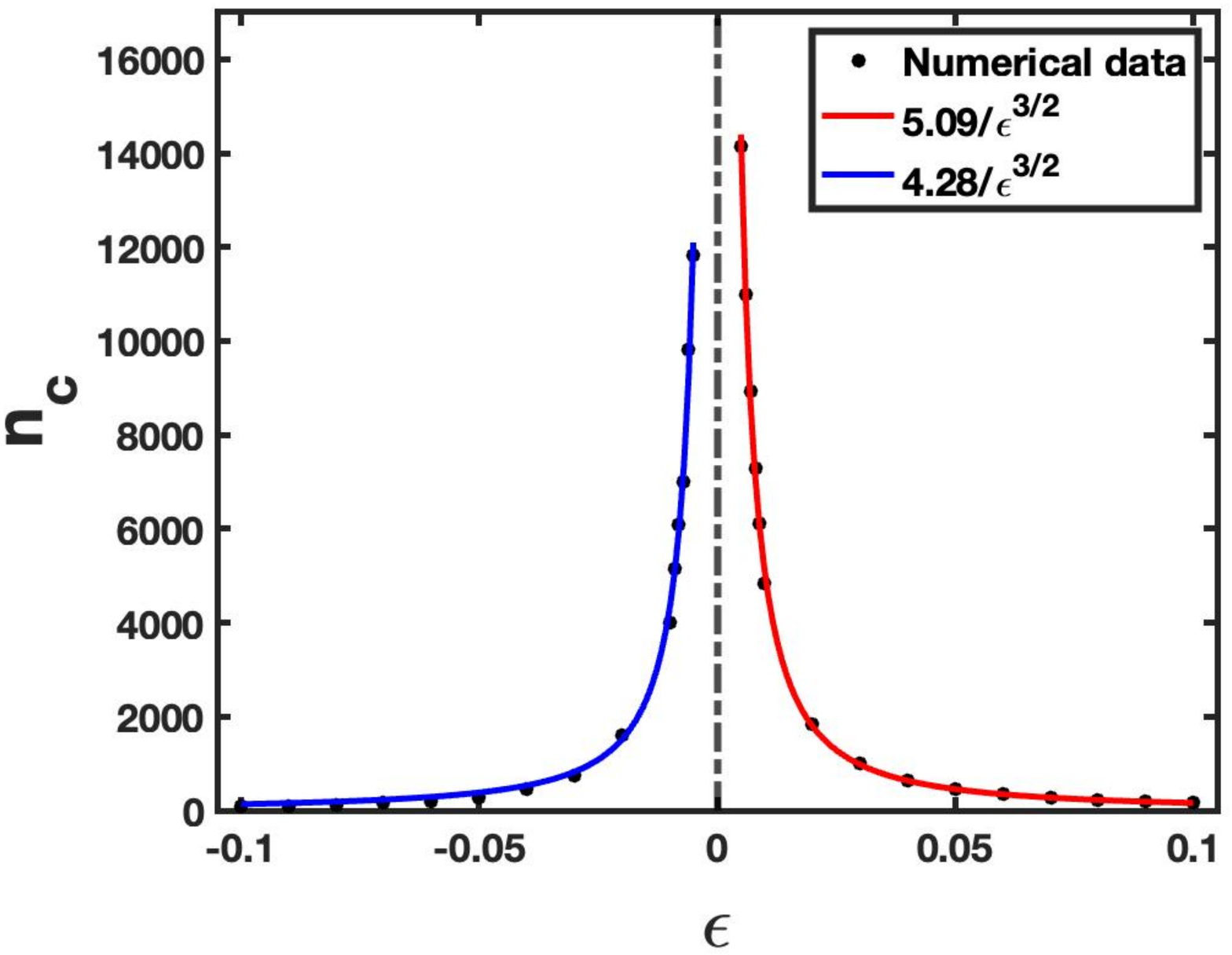}
\end{center}
\caption{Plot of crossover scale $n_c$ versus $\epsilon$, for $\ga_1
= 1, ~ \ga_2 = -1, ~a = 6$, $\om = 4 a /(\mu_1 + \ep)$, and $\phi =
\pi/4$ for the initial state. We take $n_c$ as the point where the
guiding line for $1/n^{1/2}$ behavior (black dashed line) first
crosses the numerical result in the left panel of
Fig.~\ref{sshfig02} and where the first large dip in the correlation
function appears in the middle panel of Fig.~\ref{sshfig02}. We find
that $n_c$ diverges as $1/|\epsilon|^{3/2}$ as $\epsilon \to 0$.}
\label{sshfig03}
\end{figure}

Interestingly, a different scaling of $\delta C_n$ versus $n$ arises if
$\om$ is exactly equal to $\om_p = 4 a /\mu_p$ with $\mu_p$ being
the $p$-th zero of $J_0 (z)$. (We will call the $\om_p$'s critical
frequencies, $\om_1$ being the largest such frequency). Then both
$E_k$ and its first two derivatives vanish at $k=0$ as we will now
show. For definiteness, we consider the neighborhood of $\om_1$,
namely, we take $\om = 4a/(\mu_1 + \ep)$, where $|\ep| \ll 1$. We
now expand Eq.~\eqref{ek1} around $k=0$ up to order $k^3$. Using the
property $dJ_0 (z) /dz = - J_1 (z)$ and $J_1 (\mu_1) \equiv \nu_1
\simeq 0.519$, we find that \beq E_k ~\simeq~ \nu_1 ~(-2 \ep k ~+~
\mu_1 k^3). \label{ek2} \eeq Eq.~\eqref{cn1} then gives, in the
region around $k=0$, \beq \delta C_n ~\simeq~ \frac{\cos \phi}{2 \pi} \int
dk ~{\rm Re} ~\exp [i 2n T ~\nu_1~ (-2 \ep k ~+~ \mu_1 k^3)].
\label{cn3} \eeq Defining a scaled variable $k' = k n^{1/3}$, we get
\bea \delta C_n &\simeq& \frac{\cos \phi}{2 \pi n^{1/3}} \int dk' \non \\
&& \times ~{\rm Re} ~\exp [i 2 T ~\nu_1~ (-2 \ep n^{2/3} k' ~+~
\mu_1 k'^3)]. \label{cn4} \eea We see from Eq.~\eqref{cn4} that if
$\ep = 0$, i.e., $\om = \om_1$ exactly, $\delta C_n$ will go as
$1/n^{1/3}$ times a factor which does not oscillate with $n$ at
large $n$. This can be seen in Fig.~\ref{sshfig02} (c). (If $n$ is
not very large we see some oscillations which arise due to the
stationary point at $k=\pi/2$). Further, if $\ep$ is non-zero but
small, then we will still get the $1/n^{1/3}$ scaling if $|\ep|
n^{2/3} \ll 1$ since the term of order $k'$ will dominate over the
term of order $k'^3$. But if $|\ep| n^{2/3} \gg 1$, the $k'$ term
will dominate over the $k'^3$ term, and we do not expect to get the
$1/n^{1/3}$ scaling anymore. We will then get the other scaling,
namely, an oscillating function of $n$ times $1/n^{1/2}$. Hence a
crossover will occur between a non-oscillating function of $n$ times
$1/n^{1/3}$ and an oscillating function of $n$ times $1/n^{1/2}$ at
a crossover $n_c$ which scales with $\ep$ as $1/|\ep|^{3/2}$. This
is shown in Fig.~\ref{sshfig03}. Since $|\ep| ~\sim~ |\om - \om_1|$,
we see that $n_c ~\sim~ 1/|\om - \om_1|^{3/2}$. This corresponds to
$\beta_0=3$ and $a_0=1$ (Eq.~ \eqref{cros4}).

\begin{figure}[h!]
\begin{center}
\includegraphics[height=6.2cm]{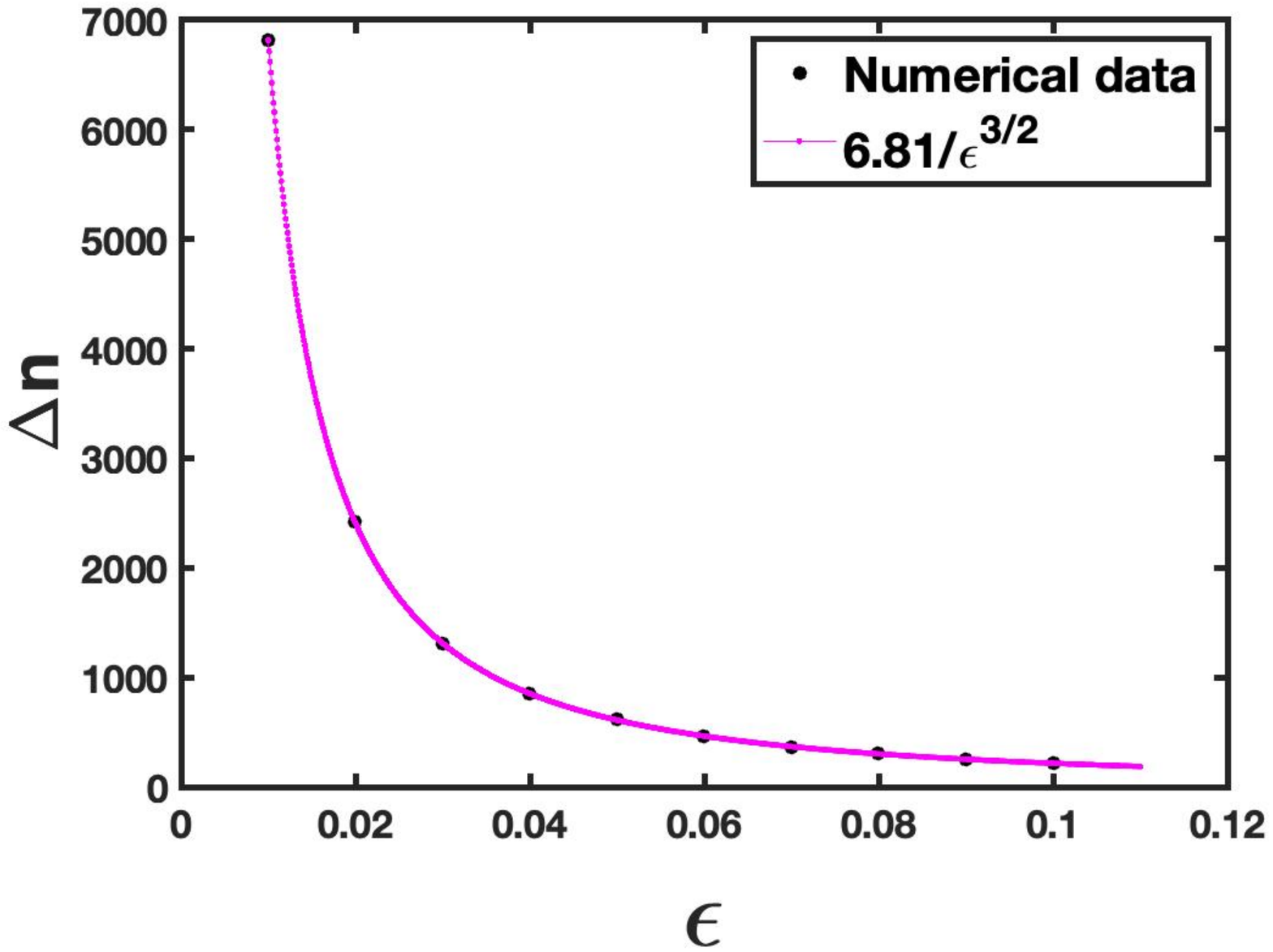}
\end{center}
\caption{Plot of oscillation period $\De n$ versus $\epsilon$, for $\ga_1
= 1, ~ \ga_2 = -1, ~a = 6$, $\om = 4 a /(\mu_1 + \ep)$, and $\phi = \pi/4$
for the initial state. We find that $\De n$ diverges as $1/|\ep|^{3/2}$
as $\ep \to 0$ from the positive side.} \label{sshfig04} \end{figure}

We now consider the oscillations which appear in $\delta C_n$ when
$n$ is larger than the crossover scale and $\om$ not equal to a
critical frequency. These are shown in Figs.~\ref{sshfig02} (a) and
(b) for $\om$ close to the value $4a/\mu_1$. For $\ep < 0$, $|\delta
C_n|$ goes as an oscillating function of $n$ times $1/n^{1/2}$ due
to the integral over the region around $k=\pi/2$; the oscillation
period is $\Delta n = \om /8$ which is independent of $\ep$ and too
small to be visible in Fig.~\ref{sshfig02} (a). But for $\ep > 0$,
we see in Fig.~\ref{sshfig02} (b) that the oscillations in $\delta
C_n$ have quite a large period, about $\Delta n = 215$. We will now
derive this. For $\ep > 0$, we see from Eq.~\eqref{ek2} that $dE_k
/dk = 0$ at $k= \pm k_0$, where \beq k_0 ~=~ \sqrt{ \frac{2 \ep}{3
\mu_1}}. \label{k0} \eeq Expanding around the stationary point at
$k_0$, we find that the argument of the exponential in
Eq.~\eqref{cn3} is given by \beq - i 2 n T ~\nu_1 ~\frac{4}{3}
~\sqrt{ \frac{2}{3 \mu_1}} ~\ep^{3/2} ~+~{\rm a ~term ~of ~order}~
(k ~-~ k_0)^2. \label{exp1} \eeq The Gaussian integral involving the
term of order $(k - k_0)^2$ will give a scaling like $1/n^{1/2}$
while the first term in Eq.~\eqref{exp1} implies that $|\delta C_n|$
will oscillate with $n$ with period \bea \Delta n &=&
\frac{\pi}{(4/3) ~2 T ~\nu_1 ~\sqrt{ \frac{2}{3 \mu_1}} ~
\ep^{3/2}} \non \\
&=& \frac{a}{(4/3) ~\mu_1 \nu_1 ~\sqrt{ \frac{2}{3 \mu_1}}
~\ep^{3/2}}, \label{deln} \eea where we have used $T = 2 \pi/\om =
\pi \mu_1/(2a)$ to derive the second line. In Fig.~\ref{sshfig04} we show
a plot of $\De n$ versus $\ep$, for $\ga_1 = 1, ~ \ga_2 = -1, ~a = 6$,
$\om = 4 a /(\mu_1 + \ep)$ for $\ep > 0$ (so that $\om < 4 a /\mu_1$), and
$\phi = \pi/4$ for the initial state. The best fit is given by $\De n =
6.81 /\ep^{3/2}$ which agrees well with the value of $6.85/\ep^{3/2}$ that
we find from Eq.~\eqref{deln}.

\begin{widetext}
\begin{figure*}[!tbp]
\includegraphics[width=0.44\hsize]{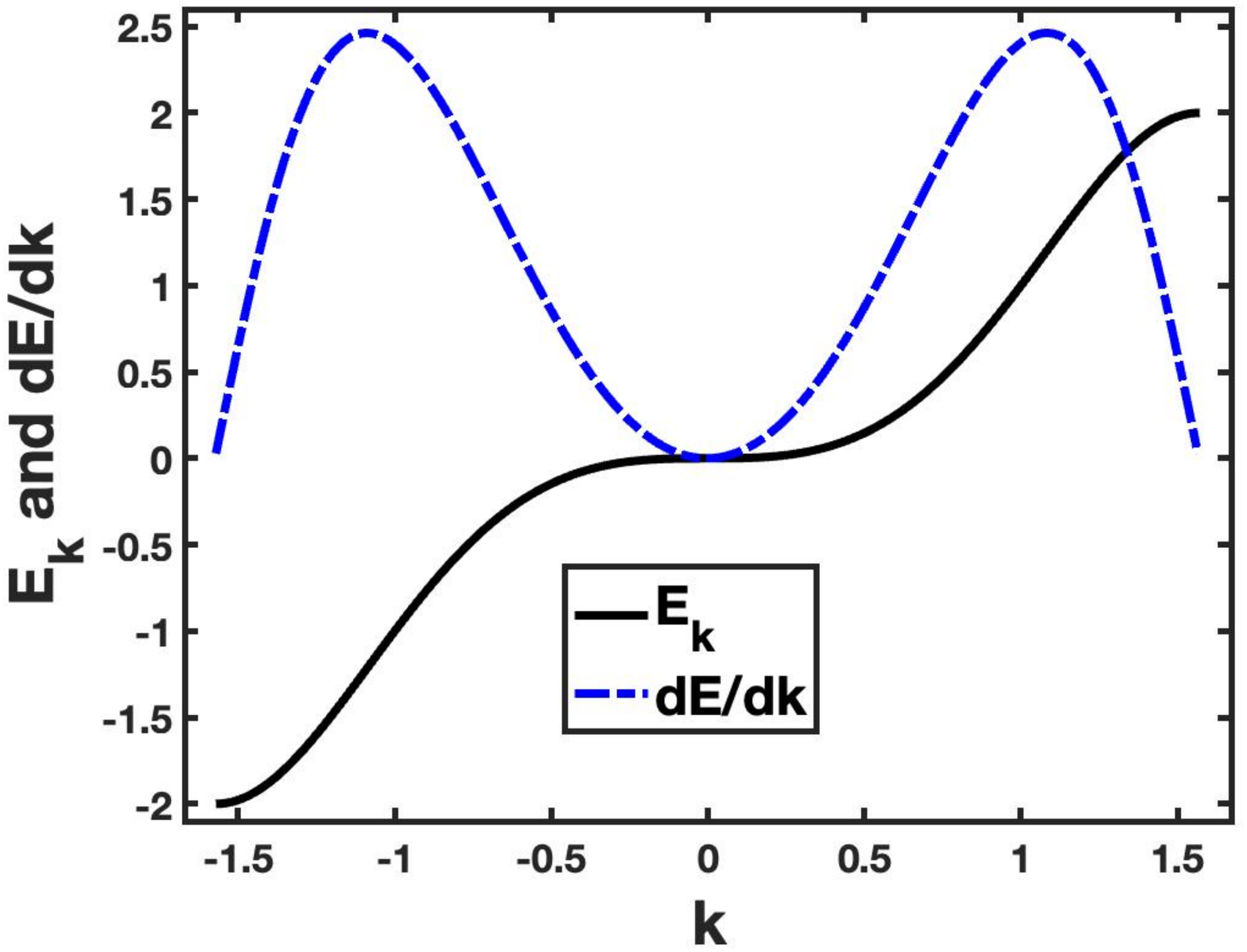}
\includegraphics[width=0.45\hsize]{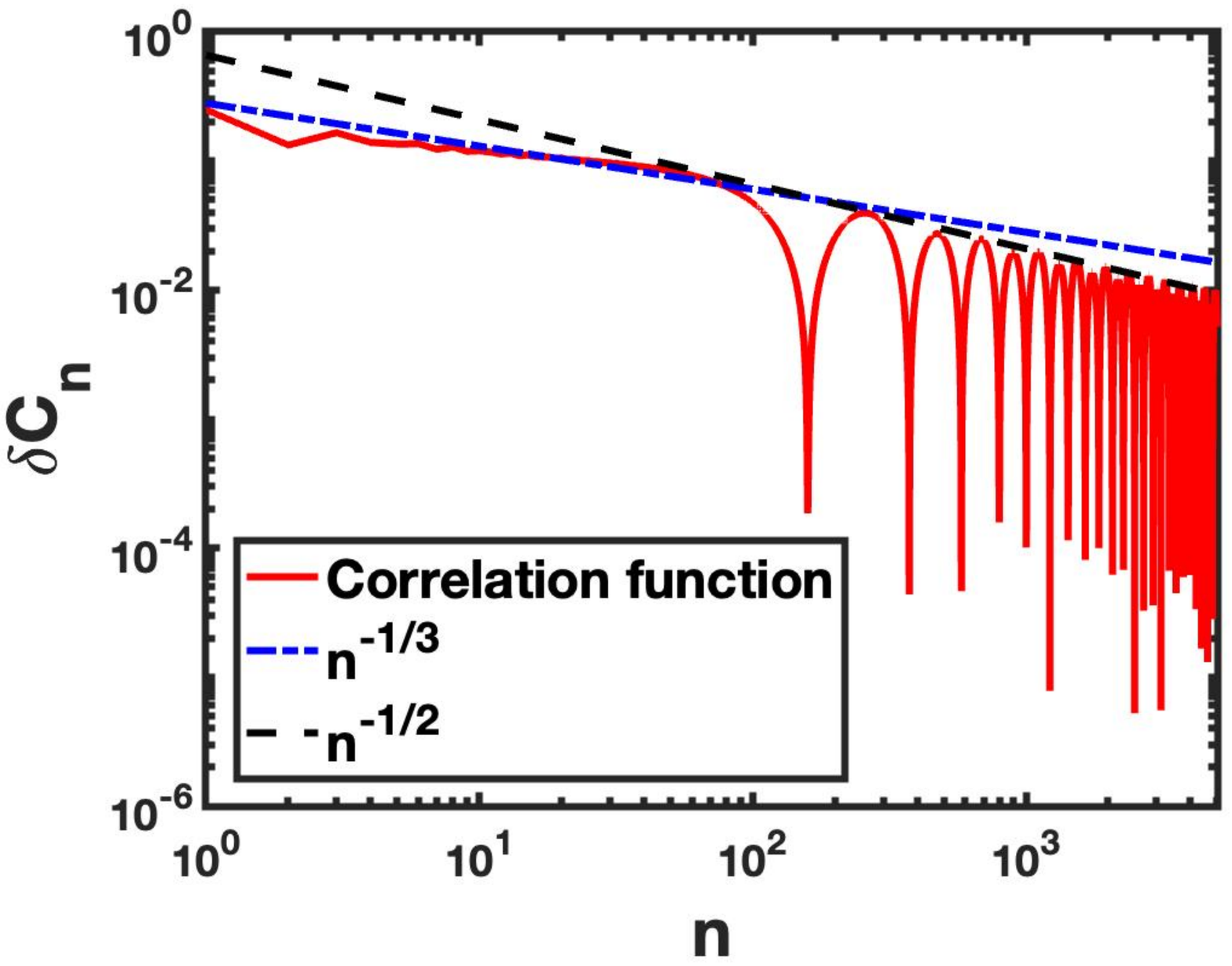}\\
\hspace*{-0.2cm} \includegraphics[width=0.45\hsize]{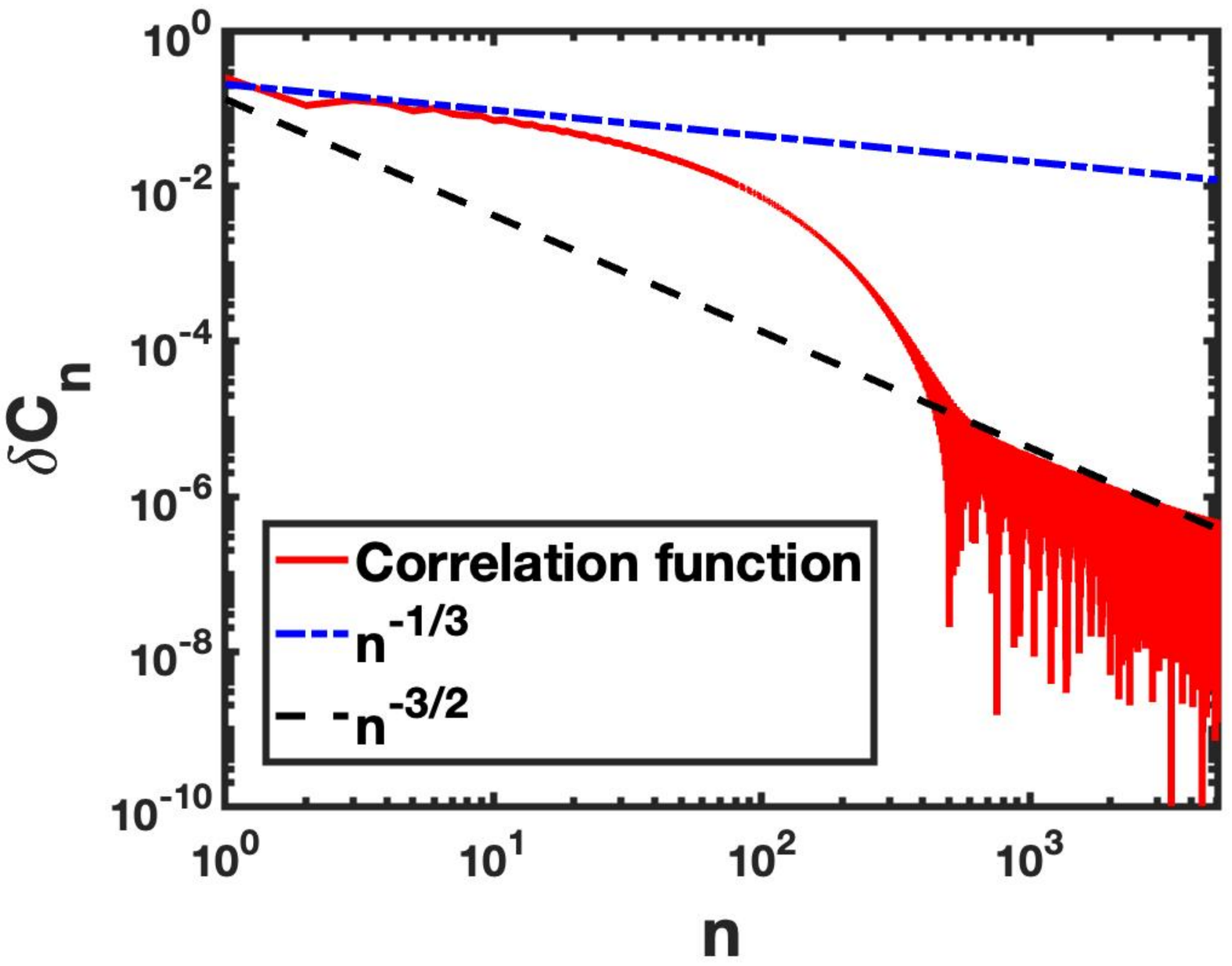}
\hspace*{0cm} \includegraphics[width=0.45\hsize]{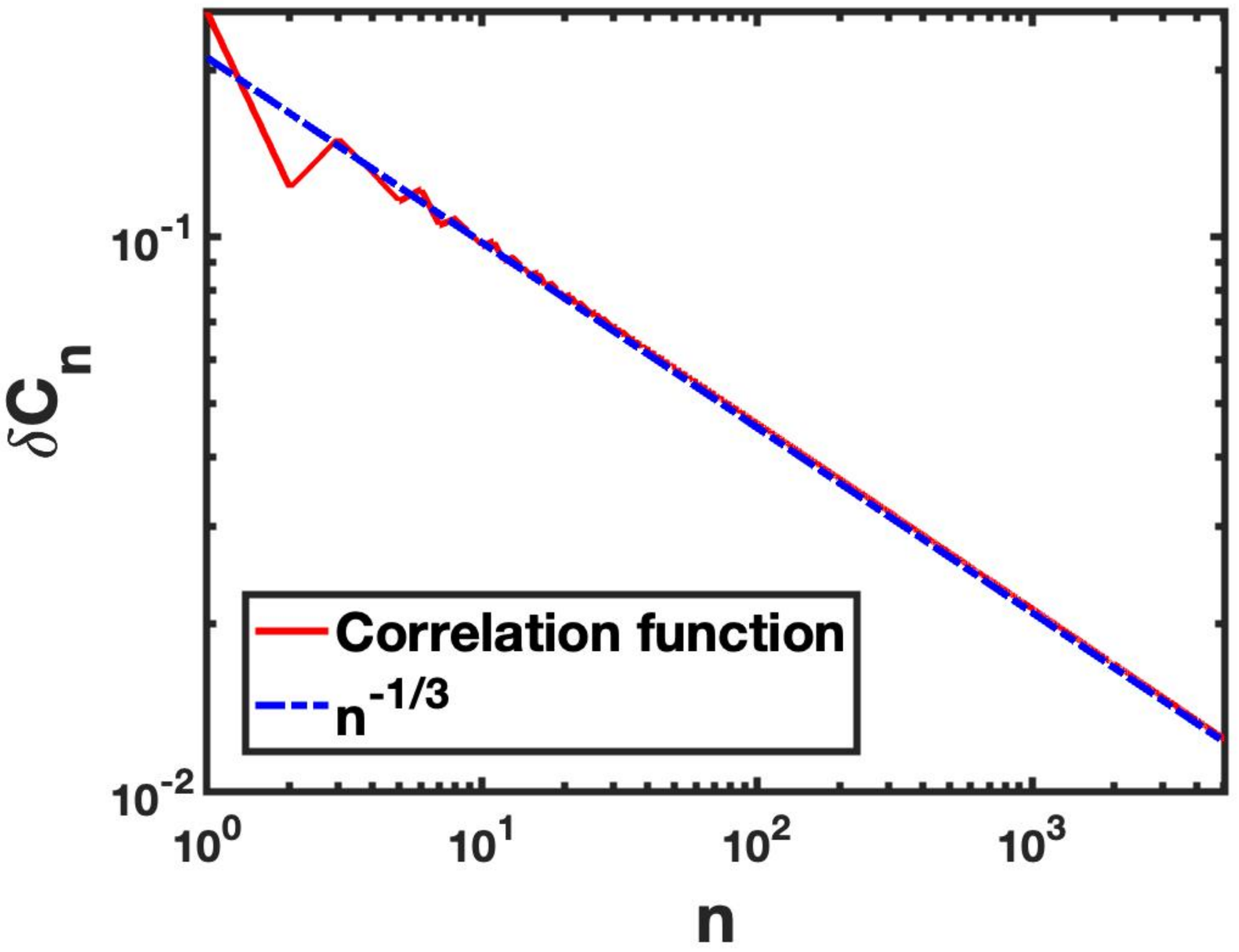}
\caption{Top left panel: Plot of the exact numerical Floquet energy
(black solid line) and its derivative (blue dot-dashed line) as a
function of $k$, for $\ga_1 = 1, ~\ga_2 = -1$ and $a=6$. Top right
panel: Log-log plot of $\delta C_n$ computed from exact numerics
(for $\phi = 0$ for the initial state) showing $n^{-1/3}$ to
$n^{-1/2}$ crossover at $\omega=4a/(\mu_1+0.1)$. Bottom left panel:
Same as top right panel but for $\omega=4a/(\mu_1-0.1)$ showing
crossover from $n^{-1/3}$ to $n^{-3/2}$ behavior. Bottom right
panel: Same as top right panel but for $\omega_c= 4a/\mu_1$ showing
$n^{-1/3}$ decay.} \label{sshfig05}
\end{figure*}
\end{widetext}

Finally, we compare our results for first-order FPT with that found
from exact numerics. The latter is shown in Fig.\ \ref{sshfig05}
for $\ga_1=-\ga_2=1$ and $a=6$. We note that these values of
$a$, $\ga_1$ and $\ga_2$, the first-order FPT yields a
critical drive frequency to be $\omega_c \simeq 9.9799$ whereas the
exact numerics leads to $\omega_c \simeq 9.9794$; this reflects the
accuracy of FPT for these parameters. (This occurs since the expansion
parameter for FPT is $\ga_1/a = 1/6$ and only odd powers of this
parameter appear. So the third-order term is about $36$ times smaller than
the first-order term). The top left panel of Fig.\
\ref{sshfig05} displays the Floquet energy and its derivative as a
function of $k$ showing that the exact Floquet energies closely
resemble the first-order theory. The top right (bottom left) panel
indicates that the crossover from $n^{-1/3}$ to $n^{-1/2}(n^{-3/2})$
behavior at $\omega= 4a/(2.4048+(-)0.1)$ is present in the exact
theory and is almost identical to that obtained within first-order
FPT. Finally, the bottom right panel shows that the $n^{-1/3}$ decay
of the correlation function at $\omega=\omega_c= 9.7994$ is
reproduced within exact numerics. The reason for this near exact
match can be traced to large value of $a$ which shifts the
transition to high frequency where first-order FPT naturally
produces accurate results.

We end this section by noting that it is not necessary for a dynamical phase
transition to have different power laws for $\om < \om_c$ and $\om > \om_c$.
We have seen above that the power law ($1/n^{1/2}$) is the same on the two
sides of $\om_c$ for a general initial state, but there is a different
power law ($1/n^{1/3}$) exactly at $\om_c$. However, for a special choice
of initial state ($\phi = 0$), the power law is different on the two sides,
being $1/n^{1/2}$ for $\om < \om_c$ and $1/n^{3/2}$ for $\om > \om_c$.

\section{Ising model}
\label{ising}

For the one-dimensional $S=1/2$ Ising model with $L$ spins and periodic
boundary conditions, the Hamiltonian reads as
\begin{eqnarray}
H = -\frac{1}{2} \sum_{j=1}^L (g\tau_j^x + \tau_j^z \tau_{j+1}^z),
\label{1dising1} \end{eqnarray}
where $\tau^{x,y,z}_j$ denote the Pauli matrices for the physical
spins on site $j$, we have set the Ising nearest-neighbor
interaction to $J=1/2$, and $g=h/J$ is the dimensionless magnetic
field. Carrying out a Jordan-Wigner transformation from spins to
spinless fermions with
\begin{eqnarray} \tau_i^x &=& 1-2c_i^\dagger c_i \nonumber \\
\tau_i^z &=& -\left[\prod_{j<i}(1-2c_j^\dagger c_j)\right](c_i^\dagger+c_i),
\label{1dising2} \end{eqnarray}
where $c_i^\dagger (c_i)$ creates (destroys) a spinless fermion on site $i$
allows one to rewrite $H$ in Eq.~\eqref{1dising1} as
\begin{eqnarray}
H &=& g \sum_{j=1}^L c_j^\dagger c_j -\sum_{j=1}^{L-1} (c^\dagger_jc_{j+1}+c^\dagger_jc^\dagger_{j+1}+\mathrm{H.c.})/2 \nonumber \\
&+& (-1)^{N_F}(c^\dagger_Lc_{1}+c^\dagger_Lc^\dagger_{1}+\mathrm{H.c.})/2,
\label{1dising3} \end{eqnarray}
where $N_F$ denotes the number of fermions. For the rest, we restrict to even
$N_F$ which implies that $c_{L+1} = -c_1$. Further using
\begin{eqnarray}
c_k =\frac{\exp(i\pi/4)}{\sqrt{L}}\sum_j \exp(-ikj)c_j, \label{1dising4}
\end{eqnarray}
where $k=2\pi m/L$ with $m=-(L-1)/2, \cdots, -1/2,1/2, \cdots, (L-1)/2$,
Eq.~\eqref{1dising3} can be written as $H=\sum_{k>0} H_k$ where
\begin{eqnarray}
H_k &=& (g-\cos k) [c_k^\dagger c_k-c_{-k}c_{-k}^\dagger] \nonumber \\
&& + \sin k [c_{-k} c_k + c_{k}^\dagger c_{-k}^\dagger]. \label{1dising5}
\end{eqnarray}
This can be recast in the form of Eq.~\eqref{hamdef} by noting that since the
fermions can be created or destroyed only in pairs, one can introduce
``pseudospins'' $|\uparrow\rangle_k =c^\dagger_k c^\dagger_{-k}|0\rangle$ and
$|\downarrow\rangle_k=|0\rangle$ where $|0\rangle$ represents the fermion
vacuum which gives
\begin{eqnarray} h_z(k,t)&=& g(t)-\cos k, \nonumber \\
h_x(k,t)&=& \sin k, \quad h_y(k,t)=0. \label{1dising6} \end{eqnarray}

We concentrate on a square pulse protocol with $g(t)=g_i$ for $0 \leq t < T/2$
and $g(t) = g_f$ for $T/2 \leq t <T$. Further, without any loss of generality,
we choose the initial state to be $(0, 1)^T$ for all $k$ which represents
the fermion vacuum or $\tau_i^x=
+1$ in terms of the physical spins to study relaxation of local quantities
to their final steady state values as a function of $n$, the number of
drive cycles. For the choice of initial state and for $L\rightarrow \infty$,
$\delta C_{ij}(n) = \langle c_i^\dagger c_j \rangle_n - \langle c_i^\dagger c_j
\rangle_\infty$ and $\delta F_{ij}(n) = \langle c_i^\dagger c_j^\dagger
\rangle_n - \langle c_i^\dagger c_j^\dagger \rangle_\infty$ equal~\cite{asen1}
\begin{eqnarray}
\delta C_{ij}(n) &=& \int_0^{\pi} \frac{dk}{2\pi} ~f_1(k) \cos (2n \phi(k)) \\
\delta F_{ij}(n) &=& \int_0^{\pi} \frac{dk}{2\pi} ~[f_2(k) \cos (2n \phi(k))
\non \\
&& ~~~~~~~~~~+ f_3(k)\sin (2n \phi(k))), \label{1dising7} \end{eqnarray}
with
\begin{eqnarray}
f_1(k) &=& -(1-\hat{n}^2_{z}(k))\cos (k(i-j)), \non \\
f_2(k) &=& -i\hat{n}_{z}(k)f_3(k) \nonumber \\
f_3(k) &=& i(n_{x}(k)+in_{y}(k)) \sin (k(i-j)). \label{1dising8} \end{eqnarray}
In Eq.~\eqref{1dising8}, we used the fact that the Floquet unitary at
each $k$ mode can be written as a $2 \times 2$ matrix of the form
$U_k = \exp[-i \phi(k) \vec{\sigma}\cdot \hat{n}(k)]$ where
$\hat{n}(k)=(n_{x}(k), n_{y}(k),n_{z}(k))$ represents a unit vector
and $\phi(k) \in [0,\pi]$ in the reduced zone scheme. The Floquet
Hamiltonian can be expressed as
\begin{eqnarray}
H_{F}(k) = \vec{\sigma}\cdot \vec{\epsilon(k)} = \Delta (k)\vec{\sigma}\cdot
\hat{n}(k)/2
\end{eqnarray}
where $\vec{\epsilon}(k) = (\epsilon_{x}(k), \epsilon_{y}(k),
\epsilon_{z}(k))$, $\Delta(k)= 2|\vec \epsilon(k)|$ is the Floquet
energy gap, and $\hat{n}(k) = \vec{\epsilon}(k) /|\vec
\epsilon(k)|$. This fixes $\phi(k) = T \Delta(k)/2$ where each
component of $\vec{\epsilon}(k)$ is restricted to $[-\pi/T,\pi/T]$
in the reduced zone scheme. The expression of $\vec \epsilon(k)$ has
been computed in Ref. \onlinecite{asen1}. For the
square pulse protocol which we focus on in this work, we find
\begin{eqnarray}
\Delta (k) &=& 2 \arccos (M_k/T), \non \\
M_k &=& \cos \Phi_i(k) \cos \Phi_f(k) \nonumber\\
&& - \hat N_i(k)\cdot \hat N_f(k) \sin \Phi_i(k) \sin \Phi_f(k), \nonumber\\
\Phi_{i(f)}(k) &=& (T/2) \sqrt{(g_{i(f)} -\cos k)^2 + \sin^2 k},
\label{deltaexp} \\
N_{i(f)}(k) &=& \left( \sin k, 0, (g_{i(f)}-\cos k) \right)T/(2
\Phi_{i(f)}(k)). \nonumber
\end{eqnarray}

The square pulse protocol allows for analytic expressions for $U_k$.
From Eq.~\eqref{1dising7}, the stationary points $d \Delta(k)/dk=0$
in $k \in [0,\pi]$ determine the behavior of the relaxation of local
quantities. As shown in Ref.~\onlinecite{asen1}, the number of
stationary points in $k \in (0, \pi)$ is $0$ for $\omega=2\pi/T
\rightarrow \infty$ while it scales as $1/\omega$ as $\omega
\rightarrow 0$. Importantly, $f_{1,2,3}(k)$ in Eq.~\eqref{1dising8}
vanish at $k=0$ and $k=\pi$ for any $(g_i, g_f, T)$ while these are
generally non-zero when $k \neq 0,\pi$. Lastly, keeping $g_i, g_f,T$
fixed, a series expansion of $\Delta(k)$ around $k=0$ and $k=\pi$
respectively yields only even powers.

For the rest, we focus on $\delta C_{ii}(n)$ which
also equals $(1-\langle \tau_i^x \rangle)/2$ from Eq.~\eqref{1dising2} (since
the initial state and the drive protocol are both translationally invariant,
the dependence on the site index $i$ can be dropped) with
the other local fermionic correlators also showing similar decays in time.
Let us quickly recapitulate the relaxation behavior in the two dynamical phases
that are distinguished by whether the stationary points occur only at
$k=0, \pi$ versus the appearance of extra stationary points in $k \in (0,\pi)$.
We denote the number of stationary points in $k \in (0,\pi)$ by $N_b$.
First, $\alpha=2$ ($\alpha=0$) for stationary points with $k =0$ or $\pi$ ($k
\neq 0, \pi$) from the behavior of $f_1(k)$. Second, $\beta=2$ in both cases.
This immediately gives a relaxation of $n^{-3/2}$ ($n^{-1/2}$)
when $N_b=0$ ($N_b \neq 0$) from Eq.~\eqref{mainr}.

\begin{widetext}
\begin{figure*}[!tbp]
\includegraphics[width=\hsize]{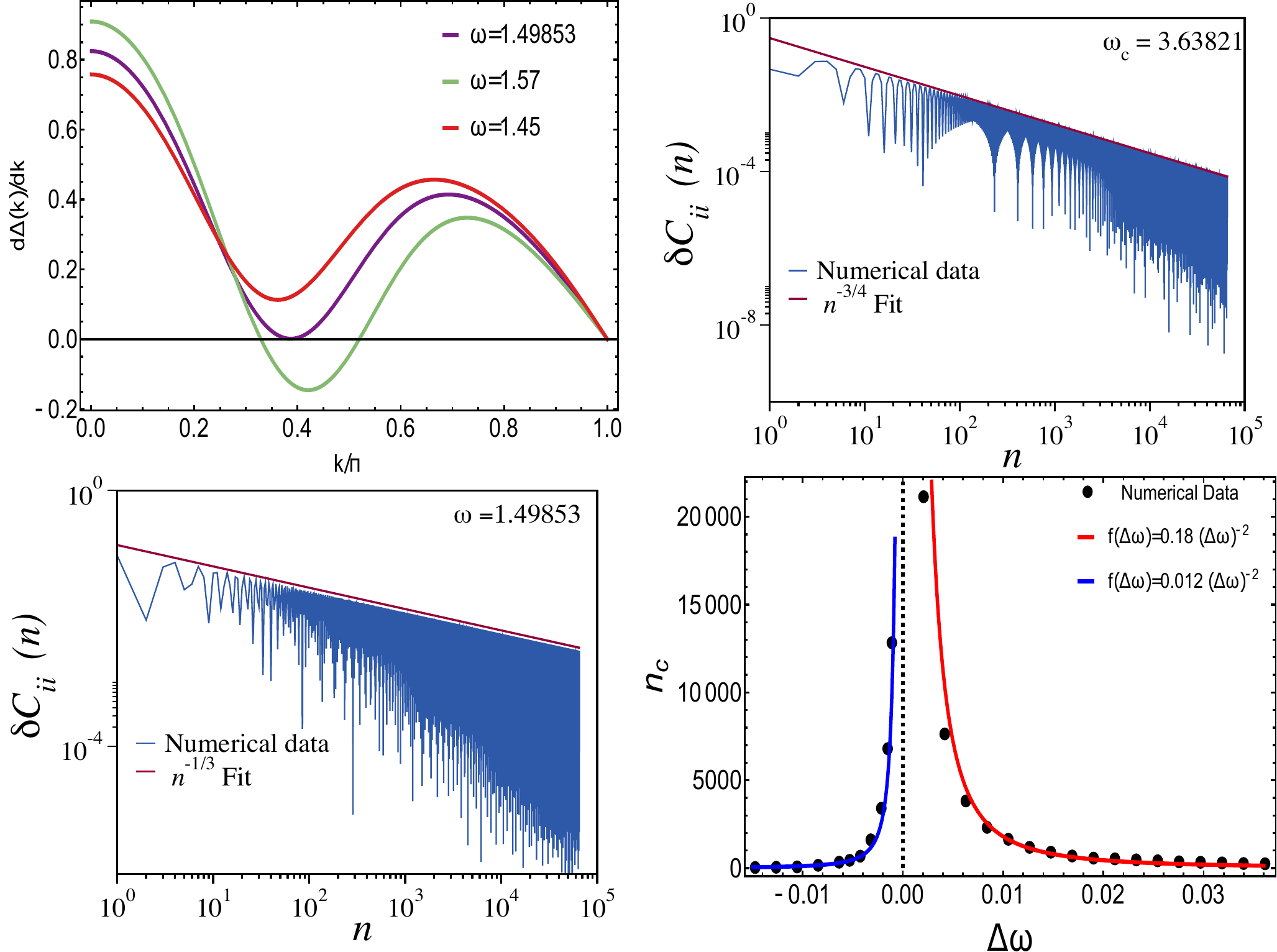}%
\caption{Top left panel: The dynamical critical point of the second
type is characterized by $N_b=1$ with this stationary point also
being a minimum of $d\Delta(k)/dk$ as the drive frequency is tuned.
Top right and Bottom left panels: Relaxation of a local quantity
$\delta C_{ii}(n)$ shown as a function of $n$ for a dynamical
critical point of the first type and of the second type (bottom
left) respectively. Bottom right panel: The behavior of the
crossover timescale, $n_c$, as the drive frequency $\omega$ is tuned
to $\omega_c \approx 3.63821$ from both sides.} \label{figising1}
\end{figure*}
\end{widetext}

We now focus on the relaxation behavior exactly at the dynamical
critical points. As discussed in Ref.~\onlinecite{asen1}, these come
in two varieties -- critical points where $N_b$ changes by $1$ (e.g.,
from $N_b=0$ to $N_b=1$) and critical points where $N_b$ changes by
two (e.g., from $N_b=2$ to $N_b=0$). The former class arises due to
an extra stationary point entering from either $k=0$ or $\pi$ and
the latter class arises due to two stationary points in $k \in
(0,\pi)$ coalescing to one at the critical point
(Fig.~\ref{figising1} (top left)) as the drive frequency is tuned
keeping $g_i, g_f$ fixed. The first dynamical transition as $\omega$
is reduced from very large values always belongs to the first category,
while some other dynamical transitions may belong to the second
category as $\omega$ as lowered further. For the first category,
while $\alpha=2$ since the extra stationary point emerges from
either $k=0$ or $\pi$, $\beta=4$ since although the critical point
requires that $d^2\Delta(k)/dk^2$ at $k=0$ or $\pi$,
$d^3\Delta(k)/dk^3=0$ at these two momenta. This implies a critical
relaxation of $n^{-3/4}$ from Eq.~\eqref{mainr}. For the second
category, the single stationary point in $k \in (0,\pi)$ also
becomes a minimum of $d\Delta(k)/dk$ (Fig.~\ref{figising1} (top
left)). Thus, $\alpha=0$ and $\beta=3$ for this point since $k \neq
0,\pi$ and $f_1(k) \neq 0$ generically for $k \in (0,\pi)$. This
gives a critical relaxation of $n^{-1/3}$ from Eq.~\eqref{mainr} for
these critical points.

\begin{widetext}
\begin{figure*}[!tbp]
\includegraphics[width=0.4\hsize]{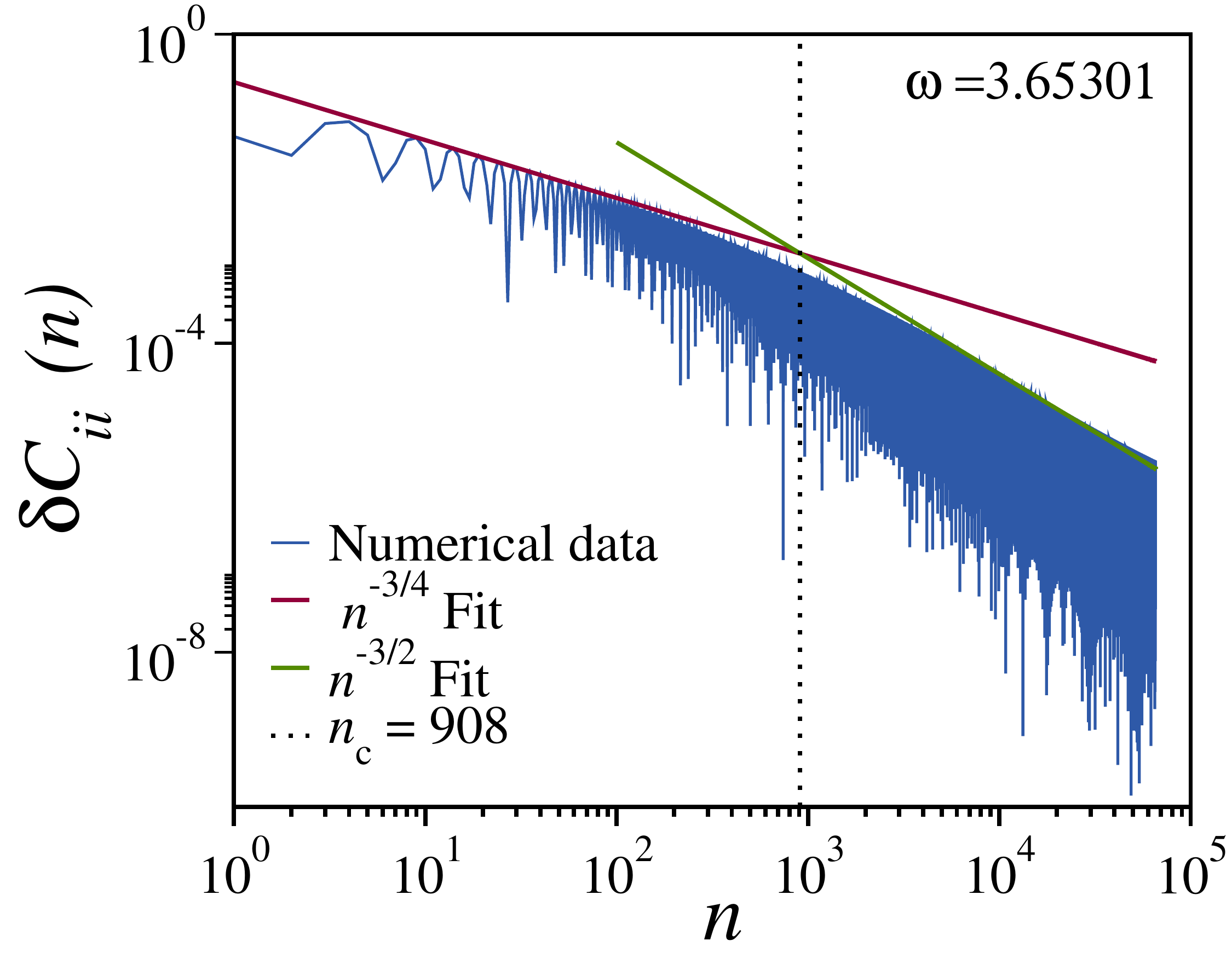}%
\includegraphics[width=0.4\hsize]{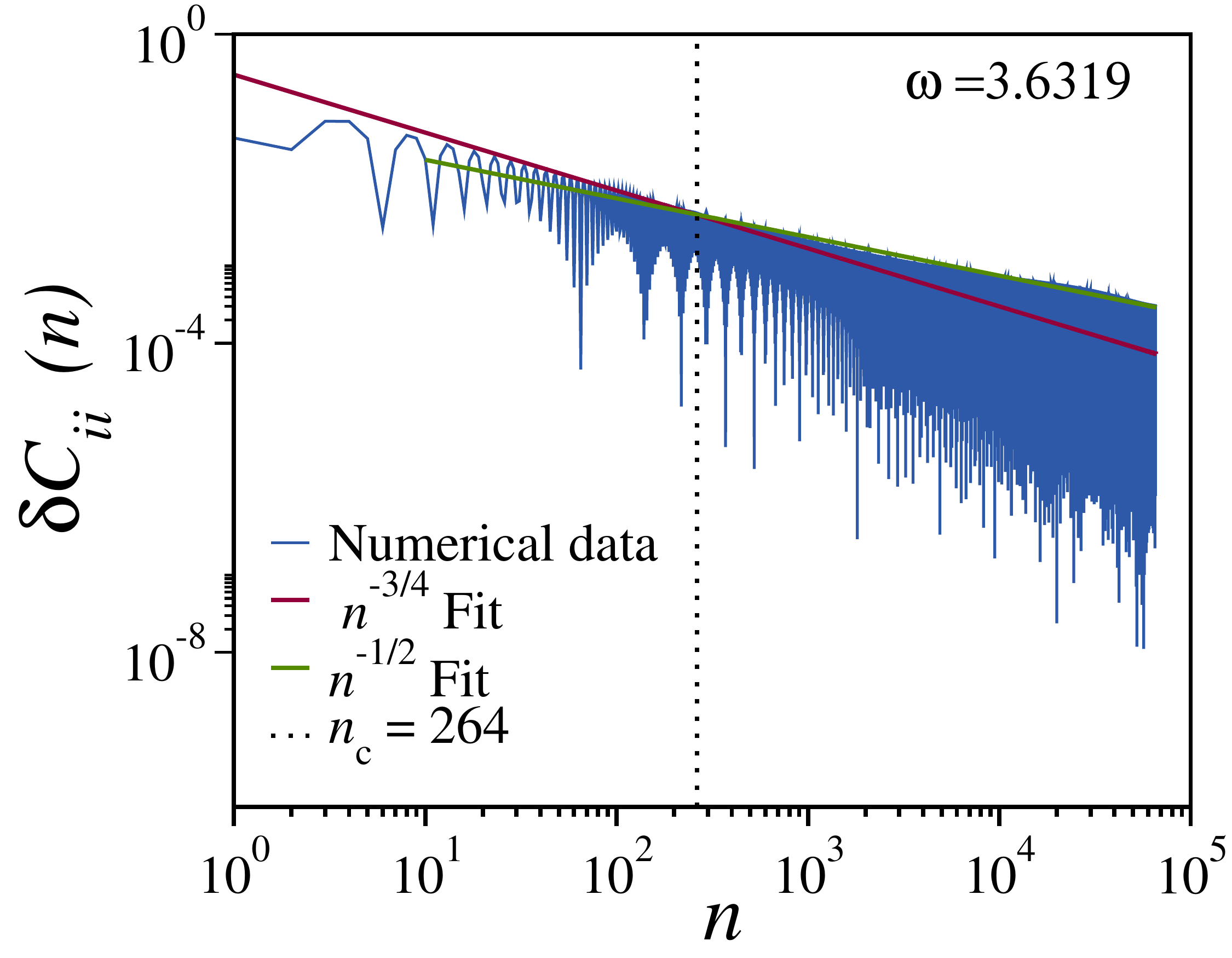}
\caption{The behavior of $\delta C_{ii}(n)$ for $\omega>\omega_c$
(left panel) and
$\omega<\omega_c$ (right panel) in the vicinity of the first dynamical
critical point ($\omega_c \approx 3.63821$) with $g_i=2, g_f=0$ shows
the presence of a dynamical crossover from critical scaling ($n^{-3/4}$)
to non-critical scaling ($n^{-3/2}$ in the left panel and $n^{-1/2}$ in the
right panel).} \label{figising2}
\end{figure*}
\end{widetext}

We now show results for $g_i=2$ and $g_f=0$ where both types of
dynamical critical points can be accessed by tuning the
drive frequency $\omega$ by using a system size of $L=8 \times 10^5$ to
minimize finite-size effects. For $\omega \approx 3.63821$, we encounter
the first dynamical critical transition where $N_b$ changes from $0$ to $1$
across the transition while for $\omega \approx 1.49853$, we encounter a
dynamical transition where $N_b$ changes from $2$ to $0$. Fig.~\ref{figising1}
(top right) shows the relaxation to be $n^{-3/4}$ for the former case and
Fig.~\ref{figising1} (bottom left) shows the relaxation to be $n^{-1/3}$ for
the latter case, completely in accord with our theoretical expectation.
Furthermore, we expect a diverging dynamical crossover timescale
$n_c$ in the vicinity
of the critical points in both the dynamical phases where the relaxation of
local quantities scale as $n^{-3/4}$ ($n^{-1/3}$) for $n \ll n_c$ before
crossing over to $n^{-3/2}$ or $n^{-1/2}$ for $n \gg n_c$. We extract $n_c$ from
our numerical data and show its behavior in the vicinity of the first dynamical
phase transition in Fig.~\ref{figising1} (bottom right). Expectedly, $n_c$
shows a divergence as the critical point is approached from both sides.
The crossover scale $n_c$ is determined by fitting the early (late) time data
for $\delta C_{ii}(n)$ to $n^{-3/4}$ ($n^{-3/2}$ or $n^{-1/2}$) and extracting
the crossing point of the fitted lines in a log-log plot
(see Fig.~\ref{figising2} (left and right panels)).

We now discuss how $n_c$ diverges near the first dynamical phase
transition as $\omega$ approaches $\omega_c$ from above. Referring
to Eq.~\eqref{cros3}, we see that here $\beta_0=4$ and $a_0=2$
since the extra stationary point enters from $k=\pi$ for the square
pulse protocol~\cite{asen1} where only even powers contribute. Thus,
$n_c \sim (c_2/c_1)^2$ and the divergence occurs since $c_1=0$
exactly at the critical point. Furthermore, $c_1$ changes sign as
$\omega$ is changed from above to below the critical frequency which
implies that $c_1 \sim \omega-\omega_c$ near the transition. This
fixes $n_c \sim (\omega-\omega_c)^{-2}$ as one approaches the
dynamical critical point from above.
\begin{widetext}
\begin{figure*}[!tbp]
\includegraphics[width=0.45\hsize]{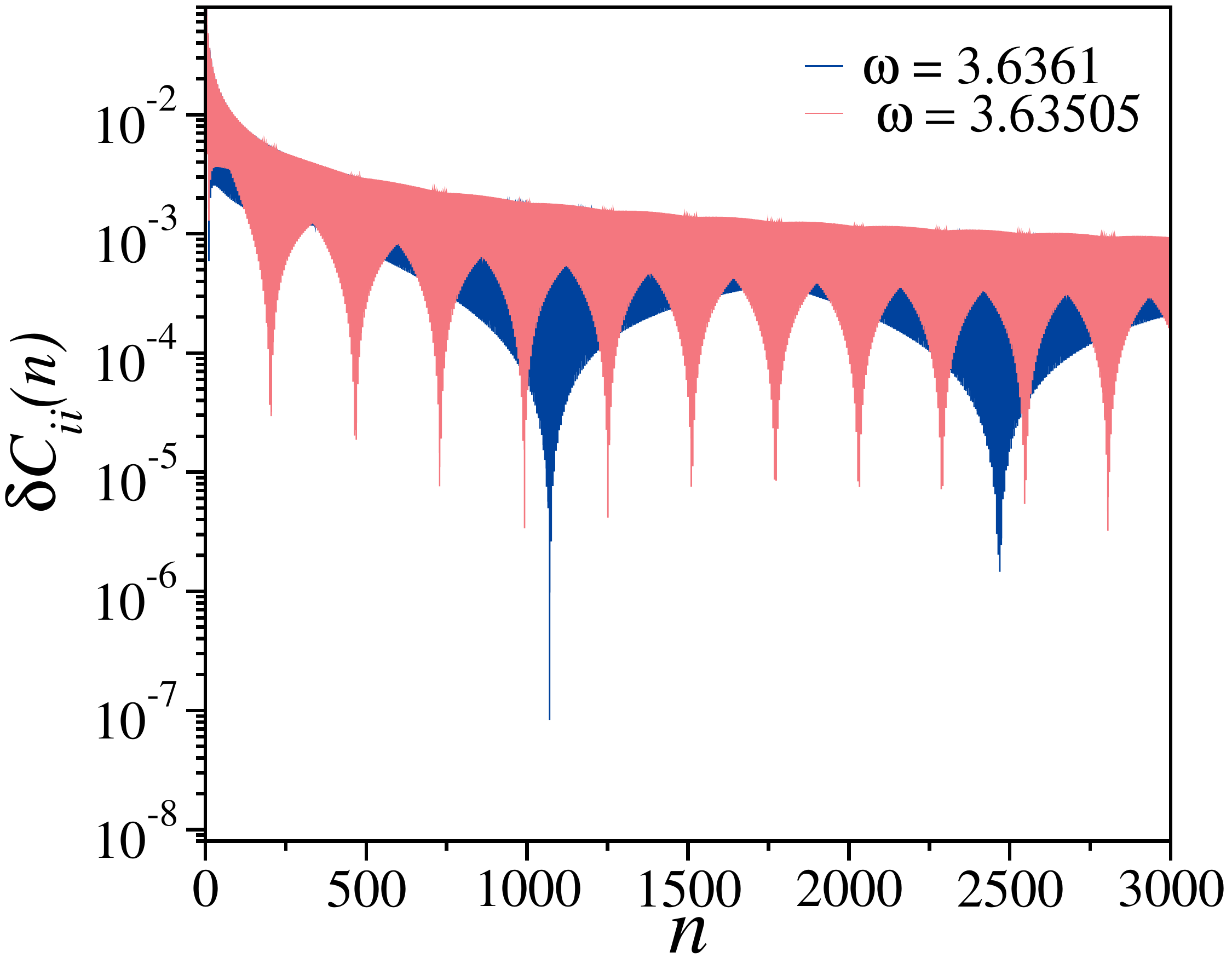} %
\includegraphics[width=0.45\hsize]{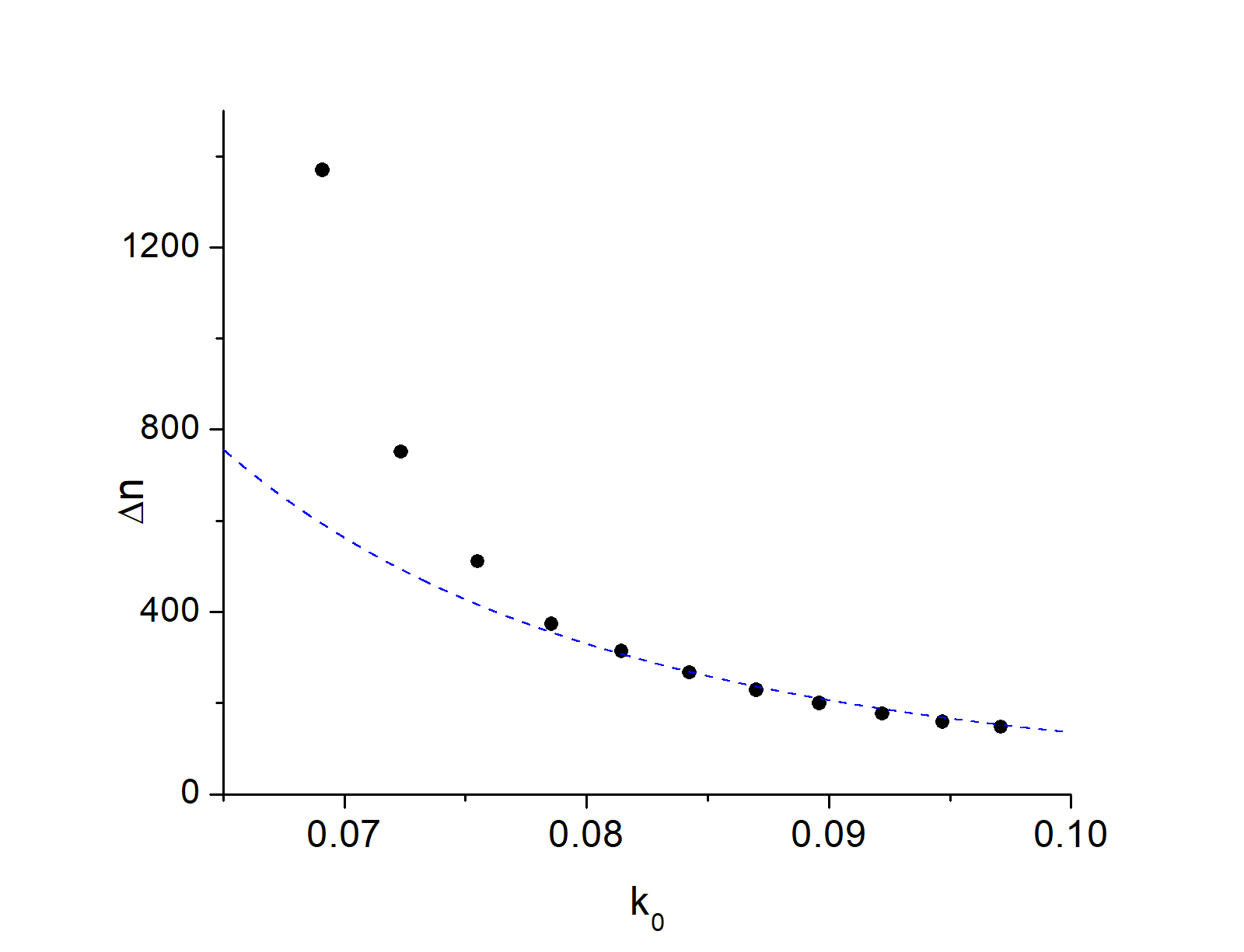}
\caption{Left panel: The behavior of $\delta C_{ii}(n)$ for two
representative values of $\omega <\omega_c \simeq 3.638$ showing
long-time coherent oscillations corresponding to $\Delta n \simeq
1400$ ($\omega=3.6361$) and $\Delta n \simeq 260$
($\omega=3.63505$). For both the plots $g_i=2$ and $g_f=0$ and
$\omega_c$ corresponds to the first transition frequency. Right
panel: A plot of the oscillation period $\Delta n$ to the distance
$k_0$ between the two extrema of $H_F$ (at $k=\pi$ and $k= \pi-k_0$)
in the Floquet Brillouin zone showing $1/k_0^4$ (the dashed blue
line corresponds to $0.0135/k_0^4$) behavior at larger $k_0$. }
\label{figosc1}
\end{figure*}
\end{widetext}
In contrast, for approaching the point from below, we need to take
into account the fact that there are two stationary points (at $k =
\pi$ and $\pi - k_0$ where $k_0 \sim \sqrt{\om_c - \om}$) which
approach each other as one nears the critical point.
Numerically, $c_1$ is small and $|c_1|$ is of the same order for
both the stationary points. The characteristic around this
stationary point controls $n_c$ and numerically we find that the
same scaling (as the one when the critical point is approached from
above) holds in this case. This is shown in the right panel of Fig.\
\ref{figising2}. A plot of the correlation function for two
representative values of $\omega <\omega_c$ is shown in the left
panel of Fig.\ \ref{figosc1}. The plot reveals a long-time
oscillation of the correlation function, similar to that identified
for the SSH model in the previous section, with $\Delta n= 1400
~(260)$ for $\omega=3.6361 ~(3.63505)$. The time period $\Delta n$
of these oscillations diverges as $\omega$ approaches $\omega_c$ in
accordance with that found for the SSH model in Sec.\ \ref{ssh}. An
analysis along the same line as in the SSH model predicts $1/k_0^4$
divergence, where $k_0 \sim \sqrt{|\omega-\omega_c|}$ is the
distance between the extrema (at $k= \pi$ and $k=\pi-k_0$) in the
Floquet Brillouin zone. This fits the data for large $k_0$; however
it breaks down when $k_0$ is small where a much faster divergence is
encountered; this is probably due to the proximity of the two
symmetry-unrelated stationary points in the Brillouin zone as well
as the small value of $d\Delta(k)/dk$ near them. These features
probably invalidate an analysis based on the premise that the
contribution to the correlation function come only from the two
stationary points.

\section{Discussion}
\label{diss}

In this work, we have studied the dynamical relaxation of
correlation functions to their steady state values in driven 1D
integrable quantum models as a function of the number of drive cycles
$n$. We summarize the generic behavior of such relaxation by
identifying a general power law in terms of two positive integers
$\alpha$ and $\beta$. The exponents corresponding to $\beta=2$ and
different $\alpha$ characterizes different dynamical phases; this
was identified in Ref.\ \onlinecite{asen1}. Here, we find the
presence of other possible exponents characterized by $\beta=3$ and
$\beta=4$. These anomalous exponents typically occur at the
dynamical transition between two dynamical phases; however, they may
also occur at special points within a dynamical phase. We provide a
general analysis of the behavior of such correlation functions in
terms of the Floquet spectrum of the driven model and show that
their occurrence is tied to points of inflections in the Floquet
spectrum. At these points, for a Floquet spectrum which is odd under
$k \to -k$, the correlation functions decay with $\beta=3$; for an even
spectrum, we find a decay with $\beta=4$.

This analysis also points to the absence of such anomalous powers
($\beta \ne 2$) for dynamical transitions in higher dimensional
integrable models. The presence of the anomalous exponent requires
the existence of a point of inflection in the Floquet spectrum; for
$d>1$, this requires vanishing of multiple derivatives $\partial^2
\Delta/\partial k_i \partial k_j$ at such a point. Since the
transition can be reached by tuning a single parameter, namely the
drive frequency, multiple derivatives cannot generically vanish at
the transition. Thus we expect such anomalous exponents to be
realized only for 1D models.

We have studied two concrete models to show the existence of such
anomalous decay. The first one involves the SSH model driven by a
continuous protocol; this model realizes decay of correlations with
$\beta=3$ leading to a $n^{-1/3}$ behavior. We analyze the driven
SSH model within first-order FPT to gain analytical insight into the
problem; the results of the first-order FPT agrees almost
identically with the exact numerical study. We also study the
correlation functions of the 1D transverse field Ising model. The
model shows a reentrant transition between two dynamical phases at
several drive frequencies. We show that the correlation function
decays with $\beta=4$ at the first (highest frequency) transition
leading to a $n^{-3/4}$ behavior. In contrast, the subsequent
transitions at lower drive frequency exhibit $n^{-1/3}$ decay and
correspond to $\beta=3$.

Near these transitions which host relaxation with anomalous power
laws, we find a crossover scale, $n_c$, after which the correlators
decay to their steady state values with exponents corresponding to
$\beta=2$. Such crossover scales can be identified at both sides of
the transition. It was found that $n_c \sim (\omega
-\omega_c)^{-\beta_0/(\beta_0-a_0)}$; thus it exhibits a power law
divergence at the transition. This behavior has been confirmed from
exact numerics for both the Ising and the SSH model. The former
model exhibits $\beta_0=4$ and $a_0=2$ leading to $n_c \sim (\omega
-\omega_c)^{-2}$ at the first dynamical transition, while the second
model corresponds to $\beta_0=3$ and $a_0=1$ leading to $n_c \sim
|\omega-\omega_c|^{-3/2}$.

Finally, our analysis shows a long-time oscillatory behavior of the
correlation functions near the transition at $\omega_c$. Such a
behavior is seen when the transition is approached from below
$\om_c$ and is seen in both models. Our FPT analysis for the SSH
model shows that such an oscillation results from the presence of
two stationary points (at $k = \pm k_0$) and provides an analytical
estimate of the time period of such oscillations. This estimate
shows a near-exact match with results from exact numerics. However,
for the Ising model, a similar analysis fails to capture the time
period when the two stationary points are close to each other (small
$k_0$); this failure could be due to proximity of symmetry unrelated
stationary points and the flat nature of $\Delta(k)$ around $k=\pi$
near the transition. This leads to near-zero values of $d
\Delta(k)/dk$ for several values of $k$ between the two stationary
points (at $k = \pi$ and $\pi - k_0$); as a result the correlators
receive contribution from all these momenta. This may invalidate an
analysis based on contributions from only the two stationary points;
we leave a further study of this issue for future work.

In conclusion, we have studied the dynamical relaxation of
correlation function of driven 1D quantum integrable models. We have
identified anomalous power laws characterizing the decay of these
correlators to their steady state value as a function of the number
of drive cycles and a diverging crossover timescale as the dynamical
transition is approached from both sides. Our analysis also reveals
a long-time oscillatory behavior of these correlation functions near
a dynamical transition when the transition is approached from the
low-frequency side.

{\it Note added}: While this manuscript was in preparation we came
to know about a similar work unraveling anomalous power laws by
Makki, Bandopadhyay, Maity and Dutta (unpublished). Our results
agree wherever a comparison is possible.

\vspace{.8cm}
\centerline{\bf Acknowledgments}
\vspace{.5cm}

S.A. thanks MHRD, India for financial support through a PMRF.
D.S. thanks DST, India for Project No. SR/S2/JCB-44/2010 for financial support.

\end{document}